\documentclass[journal,twocolumn]{IEEEtran}
\usepackage{cite,graphicx,amssymb,amsmath,color,textcomp,array,amsthm,tabulary,tabularx}
\usepackage{extarrows,multirow,multicol}
\usepackage{bm}
\usepackage{subeqnarray}
\usepackage{float}
\usepackage{subfig}
\usepackage[noend]{algorithmic}
\usepackage{algorithm}

\newcommand{\E}{\ensuremath{\mathsf{E}}}
\newcommand{\e}[1]{\ensuremath{{\rm e}^{#1}}} 
\newcommand{\PR}{\ensuremath{\mathsf{Pr}}} 
\newcommand{\diff}{\ensuremath{{\rm d}}}
\newcommand{\erf}{\ensuremath{{\sf erf}}}
\newcommand{\Rave}{\ensuremath{{\sf R}}}

\newtheorem{lemma}{Lemma}
\newtheorem{prob}{Problem}

\newtheorem{theorem}{Theorem}

\newlength{\figwidth}
\begin{document}
\title{ 
Reconfigurable Intelligent Surface assisted Two--Way Communications:\\ Performance Analysis and Optimization
}
\author{
 Saman~Atapattu, Rongfei~Fan, Prathapasinghe~Dharmawansa, Gongpu Wang,\\ Jamie~Evans,  and Theodoros A. Tsiftsis 
\thanks{
S.~Atapattu and J.~Evans are with the Department of Electrical and Electronic Engineering,
The University of Melbourne, Victoria, Australia (e-mails: \{saman.atapattu, jse\}@unimelb.edu.au). 

 R.~Fan is with the School of Information and Electronics, Beijing Institute of Technology, Beijing 100081, P. R. China (email: fanrongfei@bit.edu.cn).
 
P.~Dharmawansa is with Department of Electronic and Telecommunications Engineering, University of Moratuwa, Moratuwa, Sri Lanka (email: prathapa@uom.lk). 

G. Wang is with the School of Computer and Information Technology, Beijing Jiaotong University, Beijing 100044, P. R. China (email: gpwang@bjtu.edu.cn). 

Theodoros A. Tsiftsis is with the Institute of Physical Internet and the School of Intelligent Systems Science and Engineering, Jinan University, Zhuhai 519070, China (e-mail: theo\_tsiftsis@jnu.edu.cn).
}
\thanks{A part of this work (outage and throughput analysis) has been accepted in the IEEE Wireless Communications and Networking Conference (WCNC) 2020 \cite{atapattu2020wcnc}}
\vspace{-1.0cm} 
}
\maketitle
\begin{abstract}
In this paper, we investigate the two-way communication between two users assisted by a re-configurable intelligent surface (RIS). The scheme that two users communicate simultaneously over Rayleigh fading channels is considered. The channels between the two users and RIS can either be reciprocal or non-reciprocal. For reciprocal channels, we determine the optimal phases at the RIS to maximize the signal-to-interference-plus-noise ratio (SINR). We then derive exact closed-form expressions for the outage probability and spectral efficiency for single-element RIS. By capitalizing the insights obtained from the single-element analysis, we introduce a gamma approximation to model the product of Rayleigh random variables which is useful for the evaluation of the performance metrics in multiple-element RIS. Asymptotic analysis shows that the outage decreases at $\left(\log(\rho)/\rho\right)^L$ rate where $L$ is the number of elements, whereas the spectral efficiency increases at $\log(\rho)$ rate at large average SINR $\rho$. For non-reciprocal channels, the minimum user SINR is targeted to be maximized. For single-element RIS, closed-form solution is derived whereas for multiple-element RIS the problem turns out to be non-convex. The latter one is solved through semidefinite programming relaxation and a proposed greedy-iterative method, which can achieve higher performance and lower computational complexity, respectively.
\end{abstract}


\vspace{-0mm}
\section{Introduction} \label{S1}
Multiple antenna systems exploit spatial diversity not only to increase throughput
but also to enhance the reliability of the wireless channel. Alternatively, radio signal propagation via man-made intelligent surfaces has emerged recently as an attractive and smart solution to replace power-hungry active components \cite{renzo2019corr}. 
Such smart radio environments, that have the ability of transmitting data without generating new radio waves but reusing the same radio waves, can thus be implemented with the aid of reflective surfaces. This novel concept utilizes electromagnetically controllable surfaces that can be
integrated into the existing infrastructure, for example, along the walls of buildings. Such a surface is frequently referred to  as Reconfigurable Intelligent Surface (RIS)
 , Large Intelligent Surface (LIS)  or Intelligent Reflective Surface (IRS)\footnote{Form now on we are going to use the term RIS.}. Its tunable and reconfigurable reflectors are made of passive or almost passive electromagnetic devices which exhibit a negligible energy consumption compared to the active elements or nodes. 
This brand-new concept has already been proposed to incorporated into various wireless techniques including multiple-input multiple-output (MIMO) systems, massive MIMO, non-orthogonal multiple access (NOMA) and backscatter communications \cite{He2019wcoml,Zheng2020coml,Zhao2020coml}. 
The RIS can make the radio environment smart by collaboratively adjusting the phase shifts of reflective elements in real time. 
Therefore, most existing work on RIS focus on phase optimization of RIS elements \cite{Wu2018gcom,Yu2019iccc,Abeywickrama2019arx,Huang2019twc,Guo2019gcom,Cui2019wcl,Shen2019coml,Di2020tvt,Wu2019icassp,Huang2020jsac}. However, there are very limited research efforts explored the communication-theoretic performance limits \cite{Han2019tvt,Nadeem2020twc,Jung2019arx,Basar2019acc,Badiu2020coml,Zhao2020wcoml}. 
\vspace{-3mm}
\subsection{Related Work}
An RIS-enhanced point-to-point multiple-input single-output (MISO) system is considered in \cite{Wu2018gcom}, which aims to maximize the total received signal power at the user by jointly optimizing the (active) transmit beamforming at the access point and (passive) reflect beamforming at RIS. 
A similar system model is also considered in \cite{Yu2019iccc} where the beamformer at the access point and the RIS phase shifts are jointly optimized to maximize the spectral efficiency. 
For a phase dependent amplitude in the reflection coefficient, in \cite{Abeywickrama2019arx}, the transmit beamforming and the RIS reflect beamforming are jointly optimized based on an alternating optimization technique to achieve a low-complex sub-optimal solution.
For downlink multi-user communication helped by RIS from a multi-antenna base station,  both the transmit power allocation and the phase shifts of the reflecting elements are designed to maximize the energy efficiency on subject to individual link budget in \cite{Huang2019twc}. 
The weighted sum-rate of all users is maximized by joint optimizing the active beamforming at the base-station and the passive beamforming at the RIS for multi-user MISO systems in \cite{Guo2019gcom}. 
Moreover, 
optimization problems on physical layer security issues and  hybrid beamforming schemes where the continuous digital beamforming at the access point or base station  and discrete reflect beamforming at the RIS 
are considered in \cite{Cui2019wcl,Shen2019coml,Di2020tvt,Wu2019icassp,Huang2020jsac}.

\begin{figure*}[!ht]
  \centering
  \includegraphics[width=0.9\textwidth]{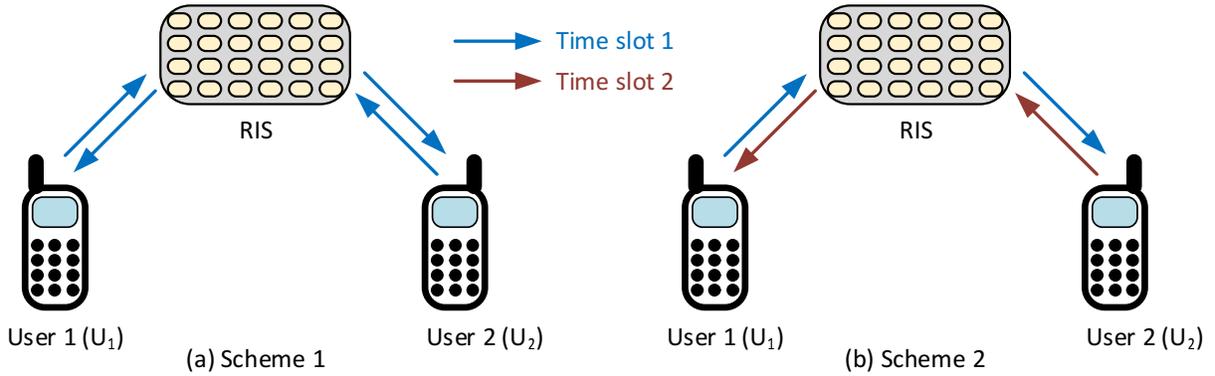}\\
  \caption{Two transmission schemes for two--way communications. (a) Scheme~1: User~1 and User~2 transmit simultaneously in time-slot~1; and (b) Scheme~2: User~1 and User~2 transmit in time-slot~1 and time-slot~2, respectively. \vspace{-5mm}
  }\label{f_scheme}
\end{figure*}

However, a few work have focused on analytical performance evaluation, and therefore very limited number of results are available so far. 
For an RIS-assisted large-scale antenna system, an upper bound on the ergodic capacity is  first derived and then a procedure for phase shift design based on the upper bound is discussed in \cite{Han2019tvt}. 
In \cite{Nadeem2020twc}, an optimal precoding strategy is proposed  when the line-of-sight (LoS) channel between the base station and the RIS is of rank-one, and some asymptotic results are also derived for the LoS channel of high-rank case. 
An asymptotic analysis of the data rate and channel hardening effect  in an RIS-based large antenna-array system is presented in \cite{Jung2019arx} where the estimation errors
and interference are taken into consideration. 
For a large RIS system, some theoretical performance limits are also explored in \cite{Basar2019acc} where the symbol error probability is derived by characterizing the receive {\sf SNR} using the central limit theorem (CLT). 
In \cite{Badiu2020coml}, the RIS transmission with phase errors is considered and the composite channel is shown to be equivalent to a point-to-point Nakagami fading channel. 

On the other hand, two--way communications exchange messages of two or more users over the same shared channel \cite{Atapattu2013tcom}, and thereby improve the spectral efficiency of the network. 
Since two--way network provides full-duplex type information exchange for the RIS networks, 
the benefits of  two--way network  are thus contingent on  proper self-interference and loop-interference  cancellations, which is possible with the recent signal processing  breakthroughs \cite{Hanzo2016proc}. 
Moreover, two--way communications have  been recently attracted considerable attention, and   have already been thoroughly investigated  with respect to most of the novel 4G and 5G wireless technologies. 
Therefore, RIS-assisted two--way networks  may also serve as a potential candidate for further performance improvement for Beyond 5G or 6G systems. However, to the best of our knowledge, all these previous  work on RIS considered the one--way communications. Motivated by this reason, as the first work, we study the RIS for two--way communications in view of quantifying the performance limits, which is the novelty of this paper.  

\vspace{-0mm}
\subsection{Summary of Contributions}


Generally speaking, although the RIS can introduce a delay, it may be negligible compared to the actual data transmission time duration.  Therefore, the transmission protocol and analytical model of the RIS-assisted two--way communication may differ from the traditional relay-assisted two--way communications \cite{Atapattu2010icst,Atapattu2013tcom}. 

Fig.~\ref{f_scheme} illustrates two possible RIS-assisted transmission schemes which require different number of time slots to achieve the bi-directional data exchange between two users.
\begin{itemize}
    \item {\bf Scheme~1 (one time-slot transmission)}: As shown in Fig.~\ref{f_scheme}\,(a), two end-users simultaneously transmit their own data  to the RIS which  reflects received signals with negligible delay. Therefore, it needs {\it only one time-slot} to exchange both users information. Since the signal is received without delay at both ends, each end-user should be implemented with {\it a pair of antennas} each for signal transmission and reception. Hence each user experiences a full-duplex type communication and {\it loop-interference and self-interference} as well.  
    \item {\bf Scheme~2 (two time-slots transmission)}: As shown in Fig.~\ref{f_scheme}\,(b),  user~1 transmits its data to  user~2 in time-slot~1, and vice versa in time-slot~2, which needs {\it two time-slots} to exchange both users information. Therefore, each end-user may use {\it a single antenna} for signal transmission and reception. Since the two users are allocated to orthogonal channels (in terms of time), they have {\it no interference} at all. This can also be interpreted as twice one-way communications. 
\end{itemize}

Since Scheme~1 is more exciting and interesting; and also Scheme~2 can be deduced from Scheme~1, we develop our analytical framework based on Scheme~1. 
\begin{figure*}[!ht]
	\centering
	\subfloat[With reciprocal channels.]{
		\label{f_system1}
		\includegraphics[width=0.45\textwidth]{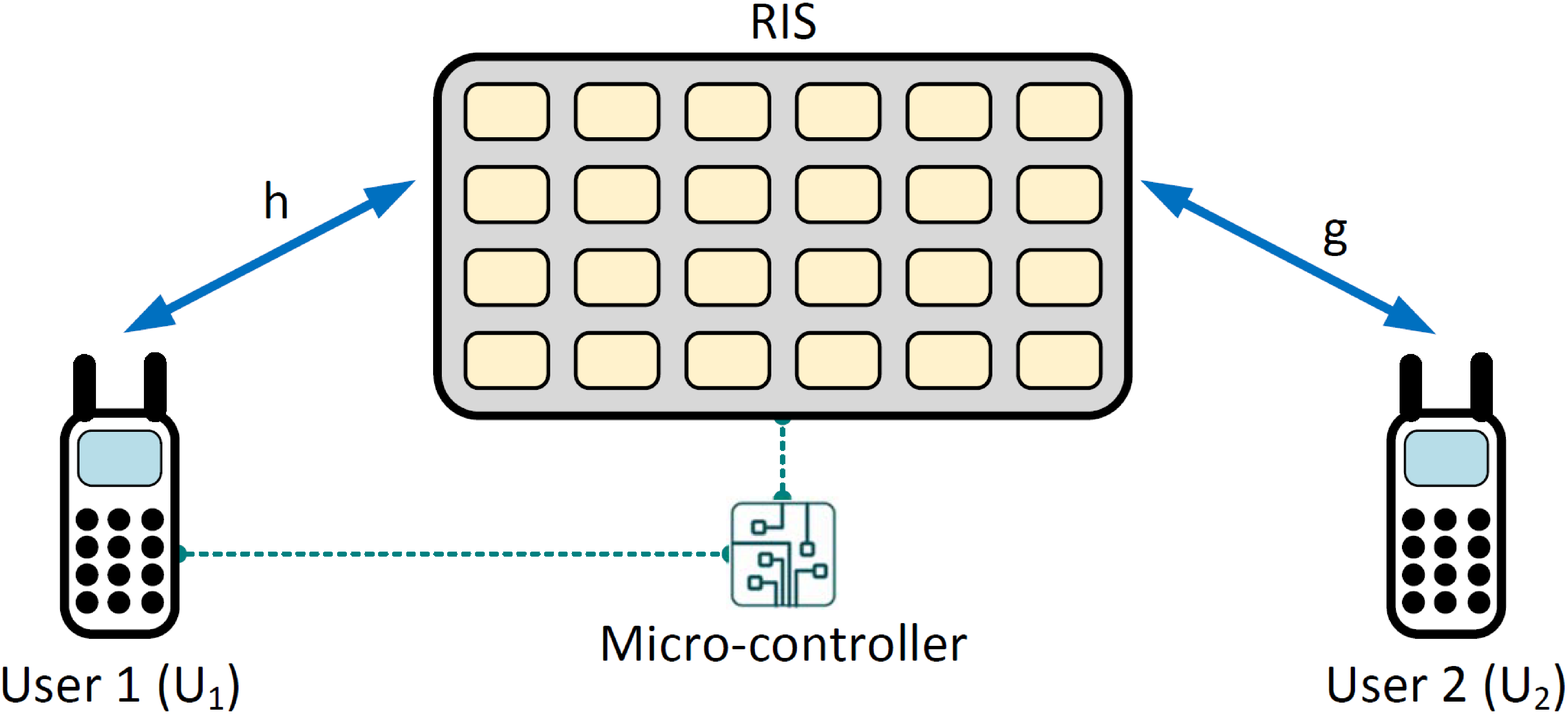}} 
        	\subfloat[With non-reciprocal channels.]{
		\label{f_system2}
		\includegraphics[width=0.45\textwidth]{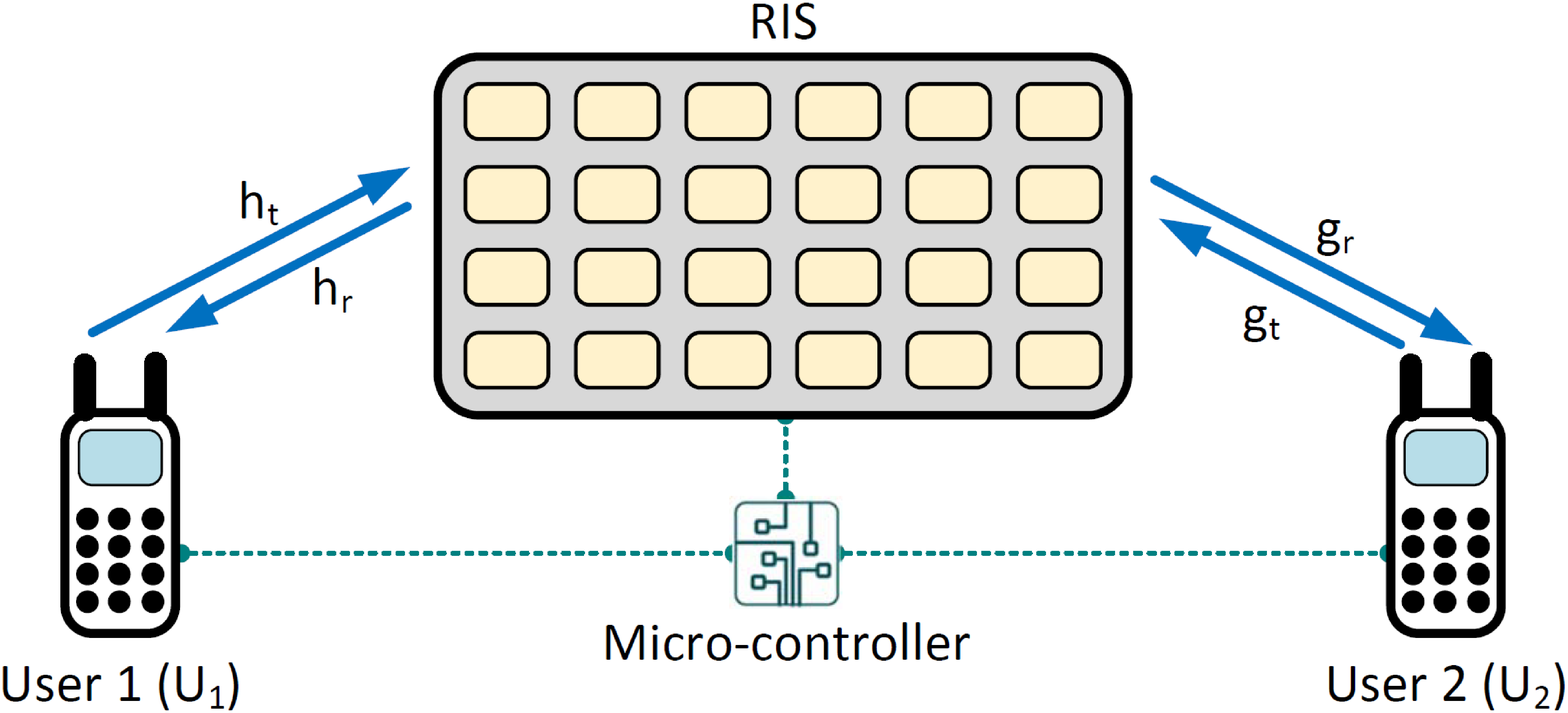}}
	\caption{Two--way communications via RIS. \vspace{-5mm}}
	\label{f_system}
\end{figure*}

Although the RIS may be implemented with large number of reflective elements for the future wireless networks, fundamental communication-theoretic foundations for single and moderate number of elements of the RIS have not been well-understood under multi-path fading. However, such knowledge is very critical for network design, e.g., distributed RIS systems.   
To support such research directions, this paper analyzes a general two--way RIS system where the number of reflective elements can range from one to any arbitrary value, and provides several communication--theoretic aspects which have not been well-understood yet. 
The main contributions of the paper are summarized as follows:

\begin{enumerate}
    
    \item {\it For reciprocal channels with a single-element RIS}, we first derive the exact outage probability and spectral efficiency in closed-form for the optimal phase adjustment at the RIS. We then provide asymptotic results for sufficiently large transmit power compared to the noise and interference powers. Our analysis reveals that the outage decreases at $\log(\rho)/\rho$ rate, whereas the spectral efficiency increases at $\log(\rho)$ rate for asymptotically large signal-to-interference-plus-noise ratio ({\sf SINR}), $\rho$.   
    
    \item {\it For reciprocal channels with a multiple-element RIS}, where the number of elements, $L$, is more than one. 
    In this respect, the  instantaneous {\sf SINR} turns out to take the form of a sum of product of two Rayleigh random variables (RVs). Since this does not admit a tractable PDF or CDF expression, we first approximate the product of two Rayleigh RVs with a Gamma RV, and then evaluate the outage probability and spectral efficiency. 
    Surprisingly, this approximation works well and more accurately than the CLT approximation (which is frequently used in LIS literature), even for a moderate number of elements such as $L=32$ or $L=64$. 
    Finally, we show that the outage decreases at $\left(\log(\rho)/\rho\right)^L$ rate, whereas the spectral efficiency still increases at $\log(\rho)$ rate.  
    
    \item 
     {\it For non-reciprocal channels}, system performance analysis seems an arduous task, since four different channel phases are involved. In this case, we turn to optimize the phase so as to maximize an important measure: the minimum user {\sf SINR}, which represents user fairness. 
    For multiple-element RIS, the associated problem is non-convex. To find the solution, through some transformations, we relax the formulated problem to be a semidefinite programming (SDP), the optimal solution of which is achievable and can further render a sub-optimal solution for our originally formulated optimization problem.
    \textcolor{black}{
    Moreover, a low-complexity method is proposed, which discretizes the searching space of each element's phase and improves elements' phases one-by-one iteratively.
    }


\end{enumerate}

Overall, this paper attempts to strike the correct balance between the performance analysis and optimization of two-way communications with the RIS.  

\emph{Notation:} Before proceeding further, here we introduce a list of symbols that have been used in the manuscript. We use lowercase and uppercase boldface letters to denote vectors and matrices respectively. A complex Gaussian random variable $X$ with zero mean and variance $\sigma^2$ is denoted by $X\sim \mathcal{CN}(\mu,\sigma^2)$, whereas a real Gaussian random variable is denoted by $X\sim \mathcal{N}(0,\sigma^2)$. The magnitude of a complex number $z$ is denoted by $|z|$ and $\E\left[\cdot\right]$ represents the mathematical expectation operator. 


\section{System Model} \label{s_system}

A RIS--aided two--way wireless network that consists of two end users (namely, $U_1$ and $U_2$) and a reflective surface ($R$) 
where the two--way networks with reciprocal and non-reciprocal channels are shown in Figs.~\ref{f_system1} and \ref{f_system2}, respectively.  The two users exchange their information symbols concurrently via the passive RIS, which only adjusts the phases of incident signals. 
The reflection configuration of the RIS is controlled by a micro-controller, which gets necessary knowledge from the users over a backhaul link as described in Section~\ref{ss_chest}.
Each user is equipped with a pair of antennas for the transmission and  reception. 
The RIS contains $L$ reconfigurable reflectors  where the $\ell$th passive element is denoted as  $I_\ell$. 
We assume that the direct link between two users is sufficiently weak to be ignored due to obstacles and/or deep fading.
For simplicity, we assume that both users use the same codebook. The unit-energy information symbols from $U_1$ and $U_2$, randomly selected from the codebook, are denoted by $s_1$ and $s_2$, respectively. 
The power budgets are $P_1$ and $P_2$ for end users $U_1$ and $U_2$, respectively. We assume that all fading channels are independent. 
By placing the antennas of users and elements of RIS sufficiently apart,  the channel gains between different antenna pairs fade more or less independently and no correlation exist. 

\subsection{Reciprocal Channels}\label{ss_system_reciprocal}
The wireless channel can be assumed to be reciprocal if the overall user-to-RIS and RIS-to-user transmission time falls within a coherence interval of the channel and the pair of antennas are placed at  sufficiently close distance. Therefore, the forward and backward channels between user and RIS are the same, see Fig.~\ref{f_system1}. 

In this case, we denote the fading coefficients from $U_1$ to the $I_\ell$ and from $U_2$ to the $I_\ell$ as $h_\ell=\alpha_\ell \e{-j\varphi_\ell}$ and $g_\ell=\beta_\ell \e{-j\psi_\ell}$, respectively. The channels are reciprocal such that the channels from the $I_\ell$ to the two end users are also $h_\ell$ and $g_\ell$, respectively. All channels are assumed to be independent and identically distributed (i.i.d.) complex Gaussian fading with zero-mean and $\sigma^2$ variance, i.e., $h_\ell,g_\ell\sim \mathcal{CN}(0,\sigma^2)$. Therefore, magnitudes of $h_\ell$ and $g_\ell$ (i.e., $\alpha_\ell$ and $\beta_\ell$) follow the Rayleigh distribution.
It is assumed that the two end users know all channel coefficients, $h_1,~...,~h_L$ and $g_1,~...,~g_L$, and the $I_\ell$ knows its own channels' phase values $\varphi_\ell$ and $\psi_\ell$. 

Each user receives a superposition of the two signals via the RIS. Thus, the receive signal at $U_1$ at time $t$ can be given as
\begin{align}
y_{1}(t)  = &\underbrace{\sqrt{P_2}\left(\sum_{\ell=1}^{L} g_\ell  \e{j\phi_\ell}h_\ell\right) s_2(t)}_{\text{Desired signal}}   + \underbrace{i_1(t)}_{\text{Loop interference}} \nonumber\\
& + \underbrace{\sqrt{P_1}\left(\sum_{\ell=1}^{L} h_\ell  \e{j\phi_\ell}h_\ell\right) s_1(t)}_{\text{Self interference}} + \underbrace{w_1(t)}_{\text{Noise}} \label{e:rx u1}
\end{align}
where $\phi_\ell$ is the adjustable phase induced by the $I_\ell$, $i_1(t)$ is the receive residual loop-interference resulting from several stages of cancellation and $w_1(t)$ is the  additive white Gaussian noise (AWGN) at $U_1$ which is  assumed to be i.i.d. with  distribution $\mathcal{CN}(0,\sigma_{w_1}^2)$. 
Further, the vectors of channel coefficients between the two users and  RIS are given as ${\bf h}=[h_1,\cdots,h_L]^{\rm T}$ and ${\bf g}=[g_1,\cdots,g_L]^{\rm T}$. The phase shifts introduced by the RIS are given by a diagonal matrix as ${\bf \Phi}={\rm diag} \left([\e{j\phi_1},\cdots,\e{j\phi_L}]\right)$. Then, we can write \eqref{e:rx u1} as 
\begin{equation}\label{e:rx u1 a}
y_{1}(t)=\sqrt{P_2}{\bf h}^{\rm T} {\bf\Phi} {\bf g} s_2(t) +  \sqrt{P_1}{\bf h}^{\rm T} {\bf\Phi} {\bf h} s_1(t) + i_1(t) + w_1(t),
\end{equation}
which shows that $U_1$ receives an observation that is a combination of the other user's symbol $s_2$ and its own symbol $s_1$. Thus, $ \sqrt{P}{\bf h}^{\rm T} {\bf\Phi} {\bf h} s_1(t)$ is the self-interference term.
Since the $U_1$ has the knowledge of  ${\bf\Phi}$, $ {\bf h}$ and $s_1$, it can completely eliminate the self-interference. Therefore, after the elimination, the received {\it instantaneous} {\sf SINR} at $U_1$ can be written as
\begin{align}\label{e:sinr u1 a}
\gamma_{1} & = \frac{\left|\sqrt{P_2}\left(\sum_{\ell=1}^{L} g_\ell  \e{j\phi_\ell}h_\ell\right) s_2(t)\right|^2}{|i_1(t)|^2 + |w_1(t)|^2}.
\end{align}
To avoid loop interference,  similar to full-duplex communications, the $U_1$ applies some   sophisticated loop interference  cancellations, which results in  residual interference. Among different models used in the literature for full-duplex communications, in this paper, we adopt the model where $i_1(t)$  is i.i.d. with zero-mean, $\sigma_{i_1}^2$ variance, additive and Gaussian, which has similar effect as  the AWGN~\cite{Atapattu2019tcom}. Further, the variance is modeled as  $\sigma_{i_1}^2=\omega P_1^{\nu}$ for $P_1\geq 1$, where the two constants, $\omega>0$ and $\nu\in[0,1]$, depend on the cancellation scheme used at the user. With the aid of \eqref{e:sinr u1 a}, the instantaneous {\sf SINR} at $U_p$ where $p=1 \text{ or }2$  can be given as
\begin{align}\label{e:sinr u1 b}
\gamma_{p} & = \rho_p\left|\sum_{\ell=1}^{L}\alpha_\ell \beta_\ell \e{j(\phi_\ell-\varphi_\ell-\psi_\ell)}\right|^2 
\end{align} 
where 
\[
\rho_p= \left\{
\begin{array}{ll}
\frac{P_2}{\sigma_{i_1}^2 + \sigma_{w_1}^2}  & p=1\text{ for }U_1  \\
\frac{P_1}{\sigma_{i_2}^2 + \sigma_{w_2}^2}  & p=2\text{ for }U_2 
\end{array}
\right.,
\]
 $\sigma_{w_2}^2$ is the noise variance and $\sigma_{i_2}^2$ is the variance of residual interference at the $U_2$. It can also be modeled as  $\sigma_{i_2}^2=\omega P_2^{\nu}$. 

\subsection{Non-Reciprocal Channels}\label{ss_system_nonreciprocal}
Even though the overall user-to-RIS and RIS-to-user transmission time falls within a coherence interval of the channel, the wireless channel can be assumed to be non-reciprocal when the pair of antennas are implemented far apart each other or non-reciprocal hardware for transmission and reception. Therefore, the forward and backward channels between user and RIS may be different, see Fig.~\ref{f_system2}. 

In this case, the fading coefficients from the transmit antenna of $U_1$ to the $I_\ell$ and from the $I_\ell$ to the receive antenna of $U_1$ are denoted as $h_{{\rm t},\ell}=\alpha_{{\rm t},\ell} \e{-j\varphi_{{\rm t},\ell}}$ and $h_{{\rm r},\ell}=\alpha_{{\rm r},\ell} \e{-j\varphi_{{\rm r},\ell}}$, where $\alpha_{{\rm t},\ell}$, $\alpha_{{\rm r},\ell}$, $\varphi_{{\rm t},\ell}$ and $\varphi_{{\rm r},\ell}$ denote amplitudes and  phases, respectively. Similarly, the respective channels associated with the $U_2$ are denoted as $g_{{\rm t},\ell}=\beta_{{\rm t},\ell} \e{-j\psi_{{\rm t},\ell}}$ and $g_{{\rm r},\ell}=\beta_{{\rm r},\ell} \e{-j\psi_{{\rm r},\ell}}$. 
All channels are assumed to be independent and identically distributed (i.i.d.) complex Gaussian with zero-mean and $\sigma^2$ variance (i.e., $h_{{\rm t},\ell},h_{{\rm r},\ell},g_{{\rm t},\ell},g_{{\rm r},\ell}\sim \mathcal{CN}(0,\sigma^2)$). 
It is assumed that the two end users have full CSI knowledge, 
i.e., ${\bf h}_{{\rm t}}=[h_{{\rm t},1},~...,~h_{{\rm t},L}]$, ${\bf h}_{{\rm r}}=[h_{{\rm r},1},~...,~h_{{\rm r},L}]$, ${\bf g}_{{\rm t}}=[g_{{\rm t},1},~...,~g_{{\rm t},L}]$ and $ {\bf g}_{{\rm r}}=[g_{{\rm r},1},~...,~g_{{\rm r},L}]$; and each $I_\ell$ element knows its own channels' phases, i.e.,  $\varphi_{{\rm t},\ell},\varphi_{{\rm r},\ell},\psi_{{\rm t},\ell}$ and $\psi_{{\rm r},\ell}$.

Thus, the receive signal at $U_1$ at time $t$ can be written as
\begin{align}
y_{1}(t)  = &\sqrt{P_2}\left(\sum_{\ell=1}^{L} h_{{\rm r},\ell}  \e{j\phi_\ell}g_{{\rm t},\ell}\right) s_2(t) + i_1(t) \nonumber\\
& + \sqrt{P_1}\left(\sum_{\ell=1}^{L} h_{{\rm r},\ell}  \e{j\phi_\ell} h_{{\rm t},\ell}\right) s_1(t) + w_1(t), \label{e:rx u1 b}
\end{align}
where $ \sqrt{P_1}\left(\sum_{\ell=1}^{L} h_{{\rm r},\ell}  \e{j\phi_\ell} h_{{\rm t},\ell}\right) s_1(t)$ denotes the self-interference, which can be eliminated due to global CSI. Subsequently, loop-interference cancellation can be applied. 
We assume the same statistical properties for loop-interference as in \eqref{e:rx u1} for comparison purposes. 
Then, 
the {\sf SINR} at $U_1$ can be written as
\begin{align}\label{e:sinr u1 c}
\gamma_{1} & = \rho_1\left|\sum_{\ell=1}^{L}\alpha_{{\rm r},\ell} \beta_{{\rm t},\ell} \e{j(\phi_\ell-\varphi_{{\rm r},\ell}-\psi_{{\rm t},\ell})}\right|^2. 
\end{align} 
By performing the similar signal processing techniques as in $U_1$, the {\sf SINR} of $U_2$ can be written as
\begin{align}\label{e:sinr u2 c}
\gamma_{2} & = \rho_2\left|\sum_{\ell=1}^{L}\beta_{{\rm r},\ell} \alpha_{{\rm t},\ell} \e{j(\phi_\ell-\psi_{{\rm r},\ell}-\varphi_{{\rm t},\ell})}\right|^2. 
\end{align} 


\subsection{Channel Estimation Procedure}\label{ss_chest}

For the channel estimate stage, we use Scheme~2 which uses one--way communications twice, where  $U_1$ communicates with $U_2$ in  the first time-slot, and vice versa in the next time-slot. Therefore, we can readily use any one--way channel estimation technique proposed in the literature, e.g., \cite{Zheng2019wcl,He2019wcoml}. Then, for reciprocal channels, both end-users have knowledge of $({\bf h},{\bf g})$; and, for non-reciprocal channels, end-users $U_1$ and $U_2$  have knowledge of $({\bf h}_{{\rm r}},{\bf g}_{{\rm t}})$ and $({\bf h}_{{\rm t}},{\bf g}_{{\rm r}})$, respectively.  
For reciprocal channels, one of the end users, say $U_1$, then provides information on the required RIS reflection configuration (i.e. $\varphi_\ell$ and $\psi_\ell$, $\forall\ell$) to
the micro-controller connected to the RIS (see Fig.~\ref{f_system1}). 
For non-reciprocal channels, both end users provide information on the required RIS reflection configuration (i.e. $(\varphi_{{\rm r},\ell},\psi_{{\rm t},\ell})$ from $U_1$ and $(\varphi_{{\rm t},\ell},\psi_{{\rm r},\ell})$ from $U_2$, $\forall\ell$) to the micro-controller connected to the RIS (see Fig.~\ref{f_system2}).
These information can be provided via a low-latency high-frequency (e.g., millimeter-wave) wireless or a separate wired backhaul link.

\section{Network With Reciprocal Channels} \label{s_per_reciprocal}

\subsection{Optimum Phase Design at RIS}\label{ss_rec_phase}

A careful inspection of the structure of $\gamma_p$ given in \eqref{e:sinr u1 b}  reveals that the optimal $\phi_\ell$, which maximizes the instantaneous {\sf SINR} of each user, admits the form 
 \begin{equation}\label{e:opt phi}
    \phi_\ell^\star=\varphi_\ell+\psi_\ell \; \text{ for }\ell=1,\cdots,L.  
 \end{equation}
 This is usually feasible at the RIS as it has the global phase information of the respective channels. Now with the aid of \eqref{e:sinr u1 b}, the maximum {\sf SINR}s at $U_p$ can be given as $\gamma_p^\star$ where
 \begin{align}\label{e:sinr u1u2}
\gamma_p^\star & = \rho_p\left(\sum_{\ell=1}^{L}\alpha_\ell \beta_\ell \right)^2 .
\end{align} 
In general, the instantaneous {\sf SINR} of each user can be written as 
\begin{align}\label{e:sinr u}
\gamma = \rho\left(\sum_{\ell=1}^{L}\zeta_\ell \right)^2 \text{ where } \zeta_\ell=\alpha_\ell \beta_\ell
\end{align} 
We define $\rho$ as the average {\sf SINR}. 
We assume 
$\sigma_{i_1}^2=\sigma_{i_2}^2=\sigma_{i}^2$ and $\sigma_{w_1}^2=\sigma_{w_2}^2=\sigma_{w}^2$.


\subsection{Outage Probability}\label{ss_rec_out}
By definition, the outage probability of each user can be expressed as $P_{\rm out} = \PR\left[\gamma \leq \gamma_{\rm th}\right]$, where $\gamma_{\rm th}$ is the {\sf SINR} threshold. This in turn gives us the important relation
\begin{align}\label{e:out def}
P_{\rm out} = F_{\gamma}\left(\gamma_{\rm th}\right),
\end{align} 
where  ${F}_{\gamma}(x)$ is the cumulative distribution function (CDF) of  $\gamma$. 
To evaluate the average spectral efficiency, we need the distributions of the RV $\gamma$. For general case, RV $\gamma$ is a summation of $L$ independent RVs each of which is a product of two independent Rayleigh RVs. Since the analysis of the general case may give rise to some technical difficulties, we evaluate the average spectral efficiency for $L=1$ and $L\geq 2$ cases separately.

\subsubsection{When $L=1$}

In this case, the instantaneous {\sf SINR} of each user is $\gamma = \rho\,\zeta_1^2= \rho\left(\alpha_1 \beta_1\right)^2 $. Since $\alpha_1$ and $\beta_1$ are identical Rayleigh RVs with parameter $\sigma$, the PDF and CDF expressions can be written as  $f_X(x)=(2x/\sigma^2)\, \e{-x^2/\sigma^2}$ and  $F_X(x)=1- \e{-x^2/\sigma^2}$, respectively. 
Since the RV $\zeta_1$ is a product of two i.i.d. Rayleigh RVs, its CDF can be derived as $F_{\zeta_1}(t)  =\PR\left(\zeta_1\leq t\right)=\PR\left(\alpha_1\leq \frac{t}{\beta_1}\right)$, from  which we obtain
\begin{align}\label{e:cdf zl}
F_{\zeta_1}(t) & = 
\int_{0}^{\infty} F_{\alpha_1}\left(\frac{t}{x}\right)f_{\beta_1}\left(x\right) \diff x 
= 1- \frac{2 t}{\sigma^2} {\sf K}_1\left(\frac{2 t}{\sigma^2}\right)
\end{align} 
where the last equality results from $\int_{0}^{\infty}{\e{-\frac{b}{4x}-ax}}\diff x=\sqrt{\frac{b}{a}}{\sf K}_1\left(\sqrt{ab}\right)$ with ${\sf K}_n\left(\cdot\right)$ denoting the modified Bessel function of the second kind \cite[eq.~3.324.1]{gradshteyn2007book}. 
For a RV $Y=aX^2$ with $a>0,~X\geq 0$, we can write its CDF as $F_Y(y)=F_X(\sqrt{y/a})$. By using this fact,  the CDF of $\gamma = \rho\,\zeta_1$ can be derived as
\begin{align}\label{e:cdf gamma L1}
F_{\gamma}(t)  = 
1-\frac{2}{\sigma^2}\sqrt{\frac{t}{\rho}}{\sf K}_1\left(\frac{2}{\sigma^2}\sqrt{\frac{t}{\rho}}\right).
\end{align} 
Thus, the outage probability can be written as 
\begin{align}\label{e:out L1}
P_{\rm out|L=1}(\gamma_{\rm th}) & = 1- \frac{2}{\sigma^2} \sqrt{\frac{\gamma_{\rm th}}{\rho}} {\sf K}_1\left(\frac{2}{\sigma^2}\sqrt{\frac{\gamma_{\rm th}}{\rho}}\right). 
\end{align} 

\subsubsection{When $L\geq 2$}
In this case, the instantaneous {\sf SINR} of each user  is given in \eqref{e:sinr u}. Let us now focus on deriving the CDF of the RV $\zeta=\sum_{\ell=1}^{L}\zeta_\ell$. However, by using the exact CDF of $\zeta_\ell$ given in \eqref{e:cdf zl}, an exact  statistical characterization of the CDF $\zeta$ seems an arduous task. To circumvent this difficulty,  we first seek an approximation for the PDF and CDF of $\zeta_\ell$.    

Among different techniques of approximating distributions~\cite{Atapattu2011twc}, the moment matching technique is a popular one. In the existing literature, the regular Gamma distribution is commonly used to approximate some complicated distributions because it has freedom of tuning two parameters: 1)  the shape parameter $k$; and 2) the scale parameter $\theta$. The mean and variance of such  Gamma distribution are $k \theta$ and $k \theta^2$, respectively. The following Lemma gives the Gamma approximation for the CDF  $F_{\zeta_1}(t)$.

\begin{lemma} \label{Lemma:SNR ordering}
The distribution of the product of two i.i.d. Rayleigh RVs with parameter $\sigma$ can be approximated with a Gamma distribution which has the CDF 
\begin{align}\label{e:cdf zl apx}
F_{\zeta_\ell}(t) & \approx \frac{1}{{\sf {\sf \Gamma}}(k)} \gamma\left(k, \frac{t}{\theta}\right)
\end{align} 
where
\begin{align*}
k=\frac{\pi ^2}{(16-\pi ^2)} \text{ and } \theta=\frac{\left(16-\pi ^2\right) \sigma^2}{4 \pi}.
\end{align*}
Further, $\gamma\left(\cdot, \cdot\right)$ is the lower incomplete gamma function~\cite{gradshteyn2007book}. Note that, by definition, the lower and upper incomplete gamma functions satisfy ${\sf \Gamma}\left(a, x\right)+\gamma\left(a, x\right)={\sf \Gamma}(a)$.
\end{lemma}

\begin{IEEEproof}
Since the first and second moments of $\zeta_\ell$ in \eqref{e:cdf zl} are  $\E\left[\zeta_\ell\right]=\pi  \sigma^2/4$ and $\E\left[\zeta_\ell^2\right]=\sigma^4$, the RV $\zeta_\ell$ has  $\pi  \sigma^2/4$  mean and $ (16-\pi^2)\sigma^4/16$ variance. 
By matching the  mean and variance of the RV $\zeta_\ell$ with the   $k \theta$ mean and $k \theta^2$ variance of   the Gamma distribution, we have \eqref{e:cdf zl apx}.
\end{IEEEproof}

\begin{figure*}
	\centering
        	\subfloat[The {\rm KL} divergence  vs $\sigma$.]{
		\label{f_kl}
		\includegraphics[width=0.45\textwidth]{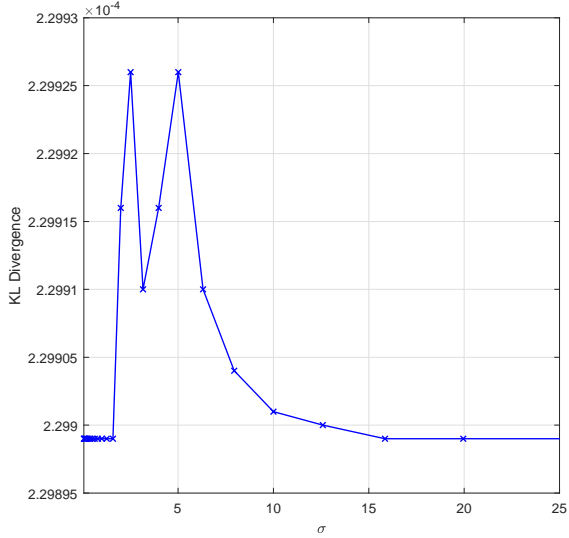}}
			\subfloat[The complementary CDF (CCDF) of $\zeta_\ell$.]{
		\label{f:ccdf ext apx}
		\includegraphics[width=0.45\textwidth]{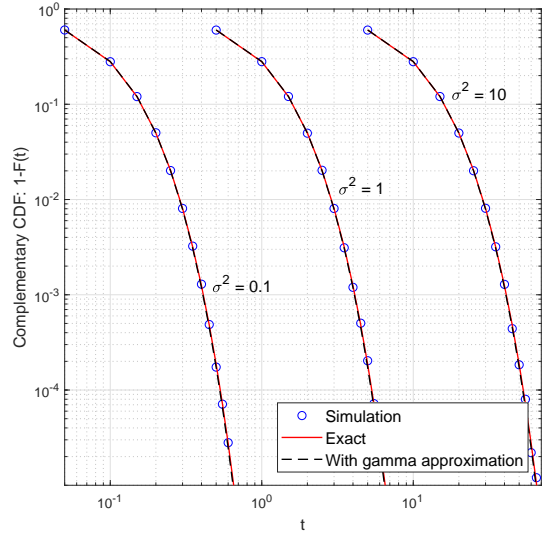}} 
	\caption{The comparison between the exact and approximation.}
	\label{f_com}\vspace{-5mm}
\end{figure*}




Here we assess the accuracy of the approximation using the Kullback-Leibler (KL) divergence. In particular, we consider the KL divergence between the exact PDF of $\zeta_\ell$ and its approximated PDF which is defined as $\mathcal D_{\rm KL}=\E\left[\log \frac{f_{\rm Ext}(t)}{f_{\rm App}(t)}\right]$ \cite{Atapattu2011twc} where the expectation is taken with respect to the exact probability density function (PDF) of $\zeta_\ell$ which can be derived as $f_{\rm Ext}(t)=4 t K_0\left(2 t/\sigma ^2\right)/\sigma ^4$. With the aid of \cite[Eq. 2.16.2.2 and 2.16.20.1]{Prudnikov83book}, we have
\begin{align}
\label{eqn KL div ray}
\mathcal D_{\rm KL}=\frac{\pi  \sigma ^2}{4 \theta }&+k \ln\left(\frac{\theta}{\sigma ^2} \right)+\epsilon  (k-2) \nonumber\\
&+\ln (4\Gamma (k)) +\E\left[\ln K_0\left(\frac{2 t}{\sigma ^2}\right)\right],
\end{align}
where $\epsilon$ is the is Euler's constant. 


 With numerical calculation, we plot the {\rm KL} divergence  vs $\sigma$ for $\sigma\in(0.05,25)$ in Fig.~\ref{f_kl}. We get $ \mathcal D_{\rm KL}\approx 2.3\times 10^{-4}$  where this very small value confirms the accuracy of the approximation. Numerical result also clarifies that $\sigma$ has a little impact on $\mathcal D_{\rm KL}$. 
Moreover, Fig.~\ref{f:ccdf ext apx} plots the complementary cumulative distribution function (CCDF) of $\zeta_\ell$ based on the simulation, the exact CDF in \eqref{e:cdf zl} and the approximate CDF in \eqref{e:cdf zl apx} for $\sigma^2=0.1,1.0,10$ which represent very small, moderate and large variance values. The exact CCDF match tightly with the Gamma approximation for the simulated $t$ range for  all $\sigma^2$, confirming the validity of the approximation. 
The accuracy of the approximation is also shown by the performance curves in Section~\ref{s:num}.

The instantaneous {\sf SINR} in \eqref{e:sinr u} admits the alternative decomposition 
\begin{align}\label{e:sinr u alt}
\gamma = \rho\,\zeta^2 
\quad{\text{where}}\quad \zeta=\sum_{\ell=1}^{L}\zeta_\ell.
\end{align} 
Armed with the above lemma, now we are in a position to derive an approximate average spectral efficiency expression pertaining to the case $L\geq 2$. It is worth mentioning here that  the RV $\zeta$ is then a sum of $L$ i.i.d. Gamma RVs with the parameters $k$ and $\theta$. Therefore, the RV $\zeta$  also follows a Gamma  distribution with  $Lk$ and $\theta$ parameters. By using the similar variable transformation as in \eqref{e:cdf gamma L1}, the CDF of $\gamma$ can be approximated as
\begin{align}\label{e:cdf gamma LL}
F_{\gamma}(t)  = 
\frac{1}{{\sf \Gamma} (L k)}{\sf \gamma} \left(L k,\frac{1}{\theta }\sqrt{\frac{t}{\rho }}\right).
\end{align} 
Therefore, the outage probability can be written as 
\begin{align}\label{e:out L2}
P_{\rm out|L\geq 2}(\gamma_{\rm th}) & \approx \frac{1}{{\sf \Gamma}\left(L k\right)} \gamma\left(L k, \frac{1}{\theta}\sqrt{\frac{\gamma_{\rm th}}{\rho}}\right) 
\end{align} 

\subsection{Spectral Efficiency}\label{ss_rec_tp}
 The spectral efficiency  can be expressed as $\log_2\left(1+ \text{\sf SINR}\right)$\,[bits/sec/Hz]. Then, the  average value  can  be evaluated as $\Rave \stackrel{}{=} \int_{0}^{\infty}\log_2\left(1+x\right)f_{\gamma}(x)\,\diff x$ where $f_{\gamma}(x)$  is the   PDF   of $\gamma$. By employing integration by parts, $\Rave$ can be evaluated as 
\begin{equation}\label{e:tp def}
\begin{split}
\Rave \stackrel{}{=} 
\frac{1}{\log(2)}\int_{0}^{\infty}\frac{1-{F}_{\gamma}(x)}{1+x}\,\diff x~\text{[bits/sec/Hz]}.  
\end{split}
\end{equation}

\subsubsection{When $L=1$}

With the aid of \eqref{e:tp def} and \eqref{e:cdf gamma L1}, the average spectral efficiency is evaluated as 
\begin{align}\label{e:tp L1}
\Rave_{\rm L=1}(\rho) & = \frac{1}{\log(2)}\frac{2}{\sigma^2\sqrt{\rho}}\int_{0}^{\infty}  \frac{\sqrt{x}}{(1+x)} {\sf K}_1\left(\frac{2}{\sigma^2}\sqrt{\frac{x}{\rho}}\right)\,\diff x \nonumber\\ & 
= \frac{1}{\log(2)\sigma^2\sqrt{\rho}} {\sf G}_{1,3}^{3,1}\left(\frac{1}{\sigma^4 \rho }\Bigg|
\begin{array}{c}
 -\frac{1}{2} \\
 -\frac{1}{2},-\frac{1}{2},\frac{1}{2} \\
\end{array}
\right)
\end{align} 
where ${\sf G}_{p,q}^{m,n}\left(\cdot \right)$ is the Meijer~${\sf G}$ function~\cite{gradshteyn2007book}. Here we have represented the Bessel function in terms of Meijer~${\sf G}$ function and subsequently use \cite[Eq. 7.811.5]{gradshteyn2007book}.

\subsubsection{When $L\geq 2$}
With the aid of \eqref{e:cdf gamma LL} and \eqref{e:tp def}, the average spectral efficiency is evaluated as 
\begin{align}\label{e:tp L2}
\Rave_{\rm L\geq 2}(\rho) & \approx \frac{1}{\log(2){\sf \Gamma} (L k)}\int_{0}^{\infty}  \frac{1}{(1+x)}{\sf \Gamma} \left(L k,\frac{1}{\theta }\sqrt{\frac{x}{\rho }}\right) \,\diff x \nonumber\\
& = \frac{1}{\log(2)} \Biggl[ 2 \log (\theta )+\log (\rho )+2 \psi ^{(0)}(L k)  \nonumber\\ 
&\quad+ \frac{\, _2{\sf F}_3\left(1,1;2,\frac{3}{2}-\frac{L k}{2},2-\frac{L k}{2};-\frac{1}{4 \theta ^2 \rho }\right)}{\theta ^2 \rho  \left(k^2 L^2-3 L k+2\right)} \nonumber \\
& \quad
+ \frac{\pi   \rho ^{-\frac{1}{2} (L k)}}{\theta ^{L k}\Gamma (L k)} 
\Biggl(
\frac{ _1{\sf F}_2\left(\frac{L k}{2};\frac{1}{2},\frac{L k}{2}+1;-\frac{1}{4 \theta ^2 \rho }\right)}{L k \left(\csc \left(\frac{\pi  L k}{2}\right)\right)^{-1}} \nonumber \\
& \quad
-\frac{ _1{\sf F}_2\Biggl(\frac{L k}{2}+\frac{1}{2};\frac{3}{2},\frac{L k}{2}+\frac{3}{2};-\frac{1}{4 \theta ^2 \rho }\Biggr)}{\sqrt{\rho } \theta(1 +  L k) \left(\sec \left(\frac{\pi  L k}{2}\right)\right)^{-1}}
\Biggr)
\Biggr]
\end{align}
where $_p{\sf F}_q\left(\cdot;\cdot;\cdot\right)$ is the generalized hypergeometric functions~\cite{gradshteyn2007book} and $\psi^{(0)}(z)$ is the logarithmic Gamma function \cite{gradshteyn2007book}. Here we have represented the Gamma function in terms of hypergeometric functions and subsequently use respective integration in \cite[Sec.~7.5]{gradshteyn2007book}.




\subsection{Asymptotic Analysis}



\subsubsection{High {\sf SINR}}
The behavior of the outage probability at high {\sf SINR} regime is given in the following theorem. 

\begin{theorem}\label{thm-1}
For high {\sf SINR}, i.e., $\rho\gg 1$, the user outage probability of $L$ elements RIS-assisted two--way networks decreases with the rate of $\left(\log (\rho)/\rho \right)^L$ over Rayleigh fading channels.  
\end{theorem}

\begin{IEEEproof}
See Appendix~\ref{App:thm-1}.
\end{IEEEproof}

However, with a traditional multiple-relay network, we observe $(1/\rho)^L$ rate. Since the end-to-end effective channel behaves as a product of two Rayleigh channels, we observe $(\log(\rho)/\rho)^L$ rate with a RIS network. This is one of the  important observations found through this analysis, and, to the best of our knowledge, this behavior has not been captured in any of the previously published work.   

The behavior of the average throughout at high {\sf SINR} regime is given in the following theorem. 

\begin{theorem}\label{thm-2}
For high {\sf SINR}, i.e., $\rho\gg 1$, the user average spectral efficiency of $L$ elements RIS-assisted two--way networks increases with the rate of $\log (\rho)$ over Rayleigh fading channels.  
\end{theorem}

\begin{IEEEproof}
See Appendix~\ref{App:thm-2}
\end{IEEEproof}

Since the residual loop-interference may also be a function of the transmit power, it is worth discussing the behavior of the outage probability and average spectral efficiency when the transmit power is relatively larger than the noise and loop interference powers. For brevity, without loss of generality, we assume $P_1=P_2=P$. The following lemmas provide important asymptotic results.   

\begin{lemma} \label{Lemma:outage}
When the transmit power is relatively larger than the noise and loop interference, i.e., $P\gg \omega,\sigma_w^2$, the outage probabilities for $L=1$ and $L\geq 2$ vary, respectively, as 
\begin{align}\label{e:out L1 p asy}
P_{\rm out|L= 1}^{\infty} & \xrightarrow{}  \left\{
\begin{array}{ll}
\frac{\gamma_{\rm th} (\omega+\sigma_w^2)}{\sigma ^4} \frac{\log (P)}{P} 
&\text{for~}\sigma_i^2=\omega  \\
P_{\rm out|L= 1}\left(\rho=\frac{1}{\omega}\right)
 & \text{for~}\sigma_i^2=\omega P 
\end{array}
\right.
\end{align} 
and 
\begin{align}\label{e:out L2 p asy}
P_{\rm out|L\geq 2}^{\infty} & \xrightarrow{}  \left\{
\begin{array}{ll}
\mathcal{G}(L,\gamma_{\rm th},\omega,\sigma) \left(\frac{\log (P)}{P} \right)^L
&\text{for~}\sigma_i^2=\omega  \\
P_{\rm out|L\geq 2}\left(\rho=\frac{1}{\omega}\right)
 & \text{for~}\sigma_i^2=\omega P 
\end{array}
\right.
\end{align} 
where $\mathcal{G}(L,\gamma_{\rm th},\omega,\sigma)\approx \frac{ \left(\gamma_{\rm th} (\sigma_w^2+\omega )\right)^{\frac{k L}{2}}}{k L \theta^{k L}\Gamma (k L)}$ is the array gain. While the outage probability decreases with the rate $\left(\frac{\log (P)}{P} \right)^L$ for $\sigma_i^2=\omega$, there is an outage floor for $\sigma_i^2=\omega\,P$. 
\end{lemma}
\begin{IEEEproof}
In particular, we consider the following two extreme cases: 
\begin{enumerate}
    \item When $\sigma_i^2=\omega$, where the interference is independent of the transmit power, we have $\rho=P/(\omega+\sigma_w^2)\xrightarrow{P\gg \omega,\sigma_w^2} \rho \propto P$. Therefore, results can easily be deduced from Theorem~\ref{thm-2}. Since we derive $L\geq 2$ case with upper and lower bounds, which are, in general, not tight. Therefore, we can derive an approximation for the array gain by using the series expansion of \eqref{e:out L2} at large $P$, which gives $\mathcal{G}(L,\gamma_{\rm th},\omega,\sigma)\approx \frac{ \left(\gamma_{\rm th} (\sigma_w^2+\omega )\right)^{\frac{k L}{2}}}{k L \theta^{k L}\Gamma (k L)}$.\footnote{It is still unclear the precise expression for the array gain $\mathcal{G}(L,\gamma_{\rm th},\omega,\sigma)$. We thus leave  it as a future work.}  
    \item When $\sigma_i^2=\omega P$, where the interference is proportional to the transmit power, we have $\rho=P/(\omega P+\sigma_w^2)\xrightarrow{P\gg \omega,\sigma_w^2} \rho \propto 1/\omega$. This means that the loop-interference variance dominates the outage probability, and respective asymptotic results can be obtained from \eqref{e:out L1} and \eqref{e:out L2} replacing $\rho$ by $1/\omega$.  
    
\end{enumerate}
This completes the proof. 
\end{IEEEproof}

When  $\sigma_i^2=\omega P^\nu$ where $\nu\in(0,1)$, it is not trivial to expand  the outage probability expressions with respect to $P$ for rational $\nu$, we omit this case. However, the performance of this case is in between $\nu=0$ and $\nu=1$ cases. 

\begin{lemma} \label{Lemma:tp}
For $P\gg \omega,\sigma_w^2$, the average spectral efficiency for $L=1$ and $L\geq 2$ vary, respectively, as
\begin{align}\label{e:tp L1 p asy}
\Rave_{L= 1}^{\infty} &  \xrightarrow{} \left\{
\begin{array}{ll}
\frac{\log (P)-\log \left(\frac{\omega+\sigma_w^2}{\sigma ^4}\right)-2 \epsilon}{\log (2)}  & \text{ for }\sigma_i^2=\omega  \\
\Rave_{L= 1}\left(\rho=\frac{1}{\omega}\right)   & \text{ for }\sigma_i^2=\omega P 
\end{array}
\right.
\end{align}
and 
\begin{align}\label{e:tp LL p asy}
\Rave_{L\geq 2}^{\infty} &  \xrightarrow{} \left\{
\begin{array}{ll}
\frac{\log (P)+2 \psi ^{(0)}(L k)-\log\left(\frac{\sigma_w^2+\omega}{\theta^2} \right)}{\log (2)}; &\hspace{-2mm}\text{}\sigma_i^2=\omega  \\
\Rave_{L\geq 2}\left(\rho=\frac{1}{\omega}\right);  &\hspace{-2mm}\text{}\sigma_i^2=\omega P 
\end{array}
\right..
\end{align}
While the average spectral efficiency increases with the rate $\log (P)$ for $\sigma_i^2=\omega$, there is a spectral efficiency floor for $\sigma_i^2=\omega\,P$.
\end{lemma}

\begin{IEEEproof}
Since the proof follows the similar steps as Lemma~\ref{Lemma:outage}, we omit the details. 
\end{IEEEproof}

Lemma~\ref{Lemma:tp} also reveals that the average spectral efficiency increases with $L$ because $\psi ^{(0)}(x)$ is an increasing function. Further, when number of elements increases from $L_1$ to $L_2(\geq L_1)$, we have $\Delta \Rave$ spectral efficiency improvements for any given $P$ where 
\begin{equation}\label{e_Rdiff}
    \Delta \Rave=\frac{2 (\psi ^{(0)}(L_2 k)-\psi ^{(0)}(L_1 k))}{\log(2) } \text{ [bits/sec/Hz]}.
\end{equation}
On the other hand, we can also save  $\Delta P$ power for any given $\Rave$ where
\begin{equation}\label{e_Pdiff}
    \Delta P=20 \log _{10}(\e{})\left(\psi ^{(0)}\left(L_2 k\right)-\psi ^{(0)}\left(L_1k\right)\right) \text{ [dBm]}.
\end{equation}
Based on the behavior of  $\psi ^{(0)}(x)$ function, the rates of $\Delta \Rave$ increment and $ \Delta P$ saving decrease with $L$. Thus, use of a very large number of elements at the RIS may not be effective compared to the required overhead cost for large number of channel estimations and phase adjustments. 

\subsubsection{For Large $L$ (or LIS)}
For a sufficiently large number $L$, according to the central limit theorem (CLT), the RV $\zeta=\sum_{\ell=1}^{L}\zeta_\ell$ converges to a Gaussian random variable with $\mu=L\pi  \sigma^2/4$  mean and $\eta=L(16-\pi^2)\sigma^4/16$ variance which has the CDF expression 
\begin{align}\label{e:cdf zL asy}
F_{\zeta}(t) & = \frac{1}{2} \left(1+\erf\left[\frac{t-\mu }{\sqrt{2 \eta} }\right]\right);\quad t\in(-\infty,+\infty)
\end{align} 
where $\erf\left[\cdot\right]$ is  the Gauss error function~\cite{gradshteyn2007book}. Since the CDF of $\gamma=\rho\,\zeta^2$ is given as $F_{\gamma}(t)=F_{\zeta}(\sqrt{t/\rho})-F_{\zeta}(-\sqrt{t/\rho})$, the outage probability can be evaluated as 
\begin{align}\label{e:out asy L2}
P_{\rm out|L\gg 1} & \approx \frac{1}{2} \left(\erf\left[\frac{\sqrt{\frac{\gamma_{\rm th}}{\rho }}-\mu }{\sqrt{2 \eta} }\right]+\erf\left[\frac{\sqrt{\frac{\gamma_{\rm th}}{\rho }}+\mu }{\sqrt{2 \eta} }\right]\right) \\ & 
= 1-{\sf Q}_{\frac{1}{2}}\left(\frac{\mu }{\sqrt{\eta} },\sqrt{\frac{\gamma_{\rm th}}{\eta\rho }}\right)
\end{align} 
where ${\sf Q}_{m}\left(\cdot,\cdot\right)$ is the Marcum's~${\sf Q}$-function and the second equality follows from the results in \cite{Annamalai2009ccnc}.

However, this CLT approximation may not be helpful to derive the average spectral efficiency in closed-form or with inbuilt special functions, which may also be a disadvantage of this approach.


\section{Network With Non-Reciprocal Channels} \label{s_nonreciprocal}

In this case, with the aid of \eqref{e:sinr u1 c} and \eqref{e:sinr u2 c}, the {\sf SINR} at $U_1$ and $U_2$ can be alternatively given as
\begin{align}\label{e:sinr u1 d}
\gamma_{1}  &= \rho_1\left|\sum_{\ell=1}^{L}c_{1,\ell}\, \e{j(\phi_\ell-\varphi_{{\rm r},\ell}-\psi_{{\rm t},\ell})}\right|^2
\text{ and }\nonumber\\
\gamma_{2}  &=  \rho_2\left|\sum_{\ell=1}^{L}c_{2,\ell}\, \e{j(\phi_\ell-\psi_{{\rm r},\ell}-\varphi_{{\rm t},\ell})}\right|^2
\end{align} 
where 
$c_{1,\ell}=\alpha_{{\rm r},\ell} \beta_{{\rm t},\ell}$ 
and $c_{2,\ell}=\beta_{{\rm r},\ell} \alpha_{{\rm t},\ell}$. 

By looking at the structures of $\gamma_{1}$ and $\gamma_{2}$, finding the optimal $\phi_\ell$, which maximizes the instantaneous {\sf SINR} of each user, is not straightforward as in the case with reciprocal channels.
This stems from the fact that the optimal $\phi_\ell$ in this case depends on phases of all channels $\varphi_{{\rm r},\ell}, \psi_{{\rm t},\ell},\psi_{{\rm r},\ell}$ and $\varphi_{{\rm t},\ell}$, and also the {\sf SINR} $\gamma_1$ is a function of $\varphi_{{\rm r},\ell}, \psi_{{\rm t},\ell}$, and  the {\sf SINR} $\gamma_2$ is a function of $\psi_{{\rm r},\ell},\varphi_{{\rm t},\ell}$. 
In this section, the optimization problem for maximizing the minimum user {\sf SINR}, i.e.,$\min(\gamma_1, \gamma_2)$, is to be formulated by optimizing the phase of the $\ell$th element of the RIS, i.e., $\phi_\ell, \forall l\in \mathcal{L}$.
We consider $L=1$ and $L\geq 2$ cases separately. 
\vspace{-0mm}
\subsection{For $L=1$}
In this case, we have 
$\gamma_{1}  = \rho_1\left|c_{1,1}\, \e{j(\phi_1-\varphi_{{\rm r},1}-\psi_{{\rm t},1})}\right|^2 = \rho_1 c_{1,1}^2 
\text{ and }
\gamma_{2}  =  \rho_2\left|c_{2,1}\, \e{j(\phi_1-\psi_{{\rm r},1}-\varphi_{{\rm t},1})}\right|^2 = \rho_2 c_{2,1}^2.$ 
Due to the fact that $|{\e{j\theta}}|=1$, the phase of each element can be any arbitrary angle. 
The outage probability and  average spectral efficiency are the same as \eqref{e:out L1} and \eqref{e:tp L1}. 
\vspace{-0mm}
\subsection{For $L\geq 2$: Problem Formulation}

For general $L$, define $\bm{\phi}=\left(\phi_1,\cdots,\phi_\ell,\cdots,\phi_L\right)^T$,
and the optimization problem that maximizes $\min(\gamma_1, \gamma_2)$ can be written as the following form
\textcolor{black}{
\vspace{0mm}
\begin{prob}\label{p:ee-1}
\begin{equation}
\begin{array}{ll}
\mathop{\max} \limits_{\bm{\phi}} \quad & \min \left(\gamma_{1},\gamma_{2} \right) \\
\end{array}
\end{equation}
\end{prob}
}
which is equivalent with the following optimization problem
\textcolor{black}{
\vspace{0mm}
\begin{prob}\label{p:ee-1-eqv}
\begin{subequations}
\begin{eqnarray}
\mathop{\max} \limits_{\bm{\phi}} \quad & t \nonumber \\
\text{s.t.} \quad & \gamma_1 \geq t \text{ and } 
            \gamma_2 \geq t. \label{e:ee-1-gamma2}
\end{eqnarray}
\end{subequations}
\end{prob}
}
Problem \ref{p:ee-1-eqv} is hard to solve directly, since both $\gamma_1$ and $\gamma_2$ are non-convex functions with $\bm{\Phi}$. 
\vspace{0mm}
\subsection{For $L\geq 2$: Solution}
To get the optimal solution of Problem \ref{p:ee-1-eqv}, we will make the following transformations.
In the first step, by resorting to (\ref{e:sinr u1 d}), 
\textcolor{black}{
we can re-write $\gamma_p$ where $p\in \{1,2\}$ as in \eqref{e_gamma1_nonrec}}, which is given on the top of this page. 
\begin{figure*}
\textcolor{black}{
\begin{equation}\label{e_gamma1_nonrec}
\begin{array}{ll}
\gamma_p & = \rho_p\left|\sum_{\ell=1}^{L}c_{p,\ell} \left( \cos\left(\phi_\ell-\varphi_{{\rm r},\ell}-\psi_{{\rm t},\ell} \right) + i \sin\left(\phi_\ell-\varphi_{{\rm r},\ell}-\psi_{{\rm t},\ell}\right) \right) \right|^2 \\
  & = \rho_p \bigg|\sum_{\ell=1}^{L}c_{p,\ell} \bigg( \cos\left(\phi_\ell\right)\cos\left(\varphi_{{\rm r},\ell}+\psi_{{\rm t},\ell}\right) + \sin\left(\phi_\ell\right)\sin\left(\varphi_{{\rm r},\ell}+\psi_{{\rm t},\ell}\right)  \\
  & \quad + i \left( \sin\left(\phi_\ell\right)\cos\left(\varphi_{{\rm r},\ell}+\psi_{{\rm t},\ell}\right) - \cos\left(\phi_\ell\right)\sin\left(\varphi_{{\rm r},\ell}+\psi_{{\rm t},\ell}\right) \right) \bigg) \bigg|^2 \\
  & = \rho_p \left(\sum_{\ell=1}^{L}c_{p,\ell} \left( \cos\left(\phi_\ell\right)\cos\left(\varphi_{{\rm r},\ell}+\psi_{{\rm t},\ell}\right) + \sin\left(\phi_\ell\right)\sin\left(\varphi_{{\rm r},\ell}+\psi_{{\rm t},\ell}\right) \right)\right)^2  \\
  & \quad + \rho_p \left(\sum_{\ell=1}^{L} c_{p,\ell} \left(\sin\left(\phi_\ell\right)\cos\left(\varphi_{{\rm r},\ell}+\psi_{{\rm t},\ell}\right) - \cos\left(\phi_\ell\right)\sin\left(\varphi_{{\rm r},\ell}+\psi_{{\rm t},\ell}\right)\right)\right)^2
\end{array}
\end{equation}
}
\hrule
\end{figure*}
For $p\in\{1,2\}$, define following $2L$-dimensional vectors:
\textcolor{black}{
\begin{align}
    \bm{\alpha}  \triangleq & \bigg(\cos(\phi_1), \sin(\phi_1), ...,  \cos(\phi_L), \sin(\phi_L)\bigg)^T, \label{e:alpha_definition} \\
    \bm{c}_p  \triangleq & \sqrt{\rho_1}\bigg(
c_{p,1} \cos\left(\varphi_{{\rm r},1}+\psi_{{\rm t},1}\right),
c_{p,1}\sin\left(\varphi_{{\rm r},1}+\psi_{{\rm t},1}\right), \nonumber \\
&\quad ...,
c_{p,L} \cos\left(\varphi_{{\rm r},L}+\psi_{{\rm t},L}\right),
c_{p,L}\sin\left(\varphi_{{\rm r},L}+\psi_{{\rm t},L}\right)
\bigg)^T,\\
\bm{d}_p \triangleq & \sqrt{\rho_1}\bigg(
-c_{p,1} \sin\left(\varphi_{{\rm r},1}+\psi_{{\rm t},1}\right),
c_{p,1} \cos\left(\varphi_{{\rm r},1}+\psi_{{\rm t},1}\right), \nonumber \\
&\quad ...,
-c_{p,L} \sin\left(\varphi_{{\rm r},L}+\psi_{{\rm t},L}\right),
c_{p,L} \cos\left(\varphi_{{\rm r},L}+\psi_{{\rm t},L}\right)\bigg)^T.
\end{align}
}

Then $\gamma_p$ where $p\in \{1,2\}$ 
can be written as 
$\gamma_p  = \bm{c}_p^T \bm{\alpha} \bm{\alpha}^T \bm{c}_p  +  \bm{d}_p^T \bm{\alpha} \bm{\alpha}^T \bm{d}_p$ which is a quadratic form of $\bm{\alpha}$. 
Define $\bm{A}= \bm{\alpha}\bm{\alpha}^T$,
$\bm{C}_p= \bm{c}_p \bm{c}_p^T $,
$\bm{D}_p= \bm{d}_p \bm{d}_p^T$, 
and 
$\bm{F}_p= \bm{C}_p+\bm{D}_p$.
It can be easily checked that the matrices $\bm{A}$, $\bm{C}_p$,  $\bm{D}_p$,  and $\bm{F}_p$ are all semi-definite positive matrices. With the above denotations,
$\gamma_p$ 
can be further written as
$\gamma_p = \text{Tr}(\left(\bm{C}_p + \bm{D}_p\right)\bm{A}) = \text{Tr}(\bm{F}_p \bm{A}) = \bm{F}_p \bullet \bm{A}$. 
For the matrix $\bm{A}$, since it is composed of $\sin(\phi_l)$ and $\cos(\phi_l)$, and $\sin(\phi_l)^2 + \cos(\phi_l)^2 = 1$ for $l\in \mathcal{L}$,  $\bm{A}$ has to satisfy the following constraint $\bm{I}_l  \bullet \bm{A}  = 1, \forall l \in \mathcal{L}$ 
where $\bm{I}_l$ is the square matrix with $(2l-1)$th and $2l$th diagonal element being 1 and all the other elements being 0.
In addition, the rank of $\bm{A}$ should be 1. Collecting the aforementioned constraints on $\bm{A}$, Problem \ref{p:ee-1-eqv} can be reformulated as the following optimization problem
\vspace{0mm}
\begin{prob} \label{p:ee-1-SDP}
\begin{subequations}
\begin{eqnarray}
\mathop{\max} \limits_{\bm{A}} \quad & t \nonumber \\
\text{s.t.} \quad &  \bm{F}_1 \bullet \bm{A} \geq t, \label{e:SDP_prob_F1} \\
             \quad &    \bm{F}_2 \bullet \bm{A} \geq t, \label{e:SDP_prob_F2} \\
             \quad &    \bm{I}_l  \bullet \bm{A} = 1, \forall l \in \mathcal{L}, \label{e: SDP_prob_sum1}\\
             \quad &  \text{Rank}(\bm{A}) = 1, \label{e:SDP_prob_rank1}\\
             \quad &  \bm{A}  \succeq 0 \label{e:SDP_prob_semidefinite}
\end{eqnarray}
\end{subequations}
\end{prob}
where $\bm{A} \succeq 0$ indicates that the matrix $\bm{A}$ is semi-definite matrix.
\textcolor{black}{
In Problem 3, (\ref{e:SDP_prob_F1}) and (\ref{e:SDP_prob_F2}) are equivalent with  (\ref{e:ee-1-gamma2}). In addition,
the constraints (\ref{e: SDP_prob_sum1}), (\ref{e:SDP_prob_rank1}) and (\ref{e:SDP_prob_semidefinite}) together guarantees that the matrix $\bf{A}$ can be decomposed to be $\bf{\alpha}\bf{\alpha}^T$ where the $\alpha$ is as defined in (\ref{e:alpha_definition}).
Collecting these facts, Problem 3 is equivalent with Problem 2.
This kind of equivalent transformation has also been broadly used in literature \cite{luo2010semidefinite}.
}

Problem \ref{p:ee-1-SDP} is also a non-convex optimization problem due to the constraint (\ref{e:SDP_prob_rank1}). 
We relax Problem \ref{p:ee-1-SDP} by dropping the constraint (\ref{e:SDP_prob_rank1}), then Problem \ref{p:ee-1-SDP}  turns to be the following optimization problem
\vspace{0mm}
\begin{prob} \label{p:ee-1-SDP_convex}
\begin{subequations}
\begin{eqnarray}
\mathop{\max} \limits_{\bm{A}} \quad & t \nonumber \\
\text{s.t.} \quad &  \bm{F}_1 \bullet \bm{A} \geq t, \label{e:a}\\
             \quad &    \bm{F}_2 \bullet \bm{A} \geq t, \label{e:b}\\
             \quad &    \bm{I}_l  \bullet \bm{A} = 1, \forall l \in \mathcal{L}, \label{e:c}\\
             \quad &  \bm{A}  \succeq 0 \label{e:d}
\end{eqnarray}
\end{subequations}
\end{prob}
For given $t$, Problem \ref{p:ee-1-SDP_convex} is a SDP feasibility problem, which can be solved with the help of CVX toolbox \cite{cvx}. 
Note that the complexity for solving Problem \ref{p:ee-1-SDP_convex} with $t$ given can be at the scale of $O(\sqrt{2L})$ according to \cite{Boyd}.
Then we need to find the maximal achievable $t$, which can be found by resorting to bisection-search method.
Denote the initial two boundary value of $t$ are $t_L$ and $t_U$ respectively, where $t_L$ makes Problem \ref{p:ee-1-SDP_convex} feasible and $t_U$ makes Problem \ref{p:ee-1-SDP_convex} infeasible.  Hence the number of iterations to achieve $\varepsilon$-tolerance, which can guarantee the difference between the searched $t$ and the maximal $t$ enabling Problem \ref{p:ee-1-SDP_convex} to be feasible lies between $\varepsilon$, is $O(\log\left(\frac{(t_U - t_L)}{\varepsilon}\right)) \sqrt{2L}$. Note that this complexity is polynomial with $L$.
\textcolor{black}{
In real application, $t_L$ and $t_U$ can be found by following Algorithm \ref{a:t_L_t_U_search}. 
\vspace{0mm}
\begin{algorithm}[H]
\caption{Search procedure for $t_L$ and $t_U$.}
\begin{algorithmic}[1] \label{a:t_L_t_U_search}
 \STATE {Initiate $i=1$ and randomly select a $t_s$ value such that $t_s>0$.}
 \WHILE {A pair of values of $t$ that make Problem \ref{p:ee-1-SDP_convex} feasible and infeasible respectively have not been found}
 \IF {Problem 4 is feasible for $t=t_s$}
 \STATE {Set $t_s = 2^i t_s$}
 \ELSE
 \STATE {Set $t_s = \frac{1}{2^i} t_s$}
 \ENDIF
 \STATE {i=i+1}
 \ENDWHILE
 \STATE {Output $t_L$ and $t_U$ as the most recent $t$ values that make Problem \ref{p:ee-1-SDP_convex}  feasible and infeasible respectively.}
  \end{algorithmic}
\end{algorithm}
\vspace{0mm}It is hard to predict the complexity of Algorithm \ref{a:t_L_t_U_search}, but with every step of $t_s$ increasing or decreasing exponentially, the convergence speed would be very fast.
}
To this end, the optimal solution of Problem \ref{p:ee-1-SDP_convex} has been found 
and the associated complexity has been characterized.

In the last step, we need to find a rank-1 solution of Problem \ref{p:ee-1-SDP_convex}. 
One broadly used method is ``Gaussian randomization procedure''  in \cite{luo2010semidefinite}. By following the idea of Gaussian random procedure, the rank-1 solution can be found in Algorithm \ref{alg:gaussian_random}.
\textcolor{black}{For brevity, the aforementioned whole procedure to solve Problem \ref{p:ee-1} when $L \geq 2$ is called as {\em SDP-relax} method.}
\begin{algorithm}[H]
\caption{Gaussian randomization procedure for Problem \ref{p:ee-1-SDP_convex}.}
\begin{algorithmic}[1]\label{alg:gaussian_random}
 \STATE {Find the optimal solution of Problem \ref{p:ee-1-SDP_convex}, which is denoted as $\bm{A}^*$ and initiate the number of randomization as $K$.}
 \FOR {k=1, 2, ..., K}
 \STATE{Generate $2L$-dimensional Gaussian random vector $\xi^k \sim\mathcal{N}(\bm{0},\bm{A}^*)$.}
 \FOR {l=1, 2, ..., L}
 \STATE {Normalize $(2l-1)$th element and $2l$th element of $\xi^k$, denoted as $\xi^k_{2l-1}$ and $\xi^k_{2l}$ by setting $\widetilde{\xi}^k_{2l-1} = \frac{\xi^k_{2l-1}}{\sqrt{\left(\xi^k_{2l-1}\right)^2 + \left(\xi^k_{2l}\right)^2}}$ and $\widetilde{\xi}^k_{2l} = \frac{\xi^k_{2l}}{\sqrt{\left(\xi^k_{2l-1}\right)^2 + \left(\xi^k_{2l}\right)^2}}$.}
 \ENDFOR
 \STATE {Generate $2L$-dimensional vector $\widetilde{\xi}^k=\left(\widetilde{\xi}_1^k, \widetilde{\xi}_2^k, ..., \widetilde{\xi}_{2L}^k\right)^T$.}
 \ENDFOR
 \STATE {Select the $k^* = \mathop{\arg\max} \limits_{k=1, 2, ..., K} \min\left(\bm{F_1} \bullet \widetilde{\xi}^k, \bm{F_2} \bullet \widetilde{\xi}^k\right)$}
\STATE  {Output $\widetilde{\xi}^{k^*}$.}
 \end{algorithmic}
\end{algorithm}

Remark: In the real application, when $K$ is larger, better solution for Problem \ref{p:ee-1-SDP_convex} can be achieved, which, however, will lead to higher computation complexity. A balanced selection of $K$ is required.


\textcolor{black}{
To further save computational complexity, a simple iterative optimization method is also proposed, which is called as {\em greedy-iterative} method for brevity. 
}
\textcolor{black}
{
In one iteration, only the phase of one element is optimized while keeping the phases of all the other elements unchanged.
The phases of multiple elements are optimized sequentially over the iterations.
In terms of optimizing the phase of one element, say $\phi_l$, the search space of $[0, 2\pi)$ is quantized into a set of discretized angles $\{0, \Delta \phi, 2\Delta \phi, ..., (K-1)\Delta \phi\}$ where $\Delta \phi = \frac{2\pi}{K}$.
With the discretization of the search space, $K$ calculations are required to  find the optimal selection of $\phi_l$ that makes $\left(\gamma_1, \gamma_2\right)$ maximal.
In this proposed method, system utility is improved in each iteration.
The searching will stop when the improvement in system utility is below a predefined threshold.
Suppose the number of iterations is $N_r$, then the total computational complexity of the proposed method is $O(N_rK)$.
One can adjust the number $K$ to find a balance between complexity and performance.
}

\vspace{0mm}
\section{Further Discussion} \label{s_diss}

\subsection{Discussion on Scheme~2}\label{ss:1wayvs2way}

While two users communicate simultaneously within a single time-slot in Scheme~1,  
$U_1$ and $U_2$ transmit the first and second time-slots, respectively, in Scheme~2. Therefore, there is no self-interference and loop-interference and this scheme can be treated as two times one--way communications.  Then, the maximum instantaneous {\sf SNR} of each user can be given with the aid of \eqref{e:sinr u} as
\begin{equation}\label{e_sch2_snr}
    \gamma = \frac{P}{\sigma_{w}^2}\left(\sum_{\ell=1}^{L}\zeta_\ell \right)^2
\end{equation}
where $P$ is the transmit power. The corresponding optimal phases are i) with reciprocal channels: $\phi_\ell^\star=\varphi_\ell+\psi_\ell$ for both users; and ii) with non-reciprocal channels: $\phi_\ell^\star=\varphi_{{\rm r},\ell}+\psi_{{\rm t},\ell}$ for $U_1$ and $\phi_\ell^\star=\varphi_{{\rm t},\ell}+\psi_{{\rm r},\ell}$ for $U_2$. 

Since there is no loop interference ($\sigma_i^2=0$) in \eqref{e_sch2_snr}, the {\sf SNR} of Scheme~2 is always larger than the {\sf SINR} of Scheme~1 for non-zero loop-interference, i.e., $\gamma=\frac{P}{\sigma_{w}^2+\sigma_i^2}\left(\sum_{\ell=1}^{L}\zeta_\ell \right)^2$ in \eqref{e:sinr u}. Therefore, Scheme~2 achieves lower outage probability which can easily be deduced from \eqref{e:out L1} and \eqref{e:out L2} replacing $\rho$ as $\rho=P/\sigma_{w}^2$. From Theorem~\ref{thm-1}, we can conclude that the user outage probability  decreases with the rate of $\left(\log (\rho)/\rho \right)^L$ over Rayleigh fading channels.

Since only one user communicates in a given frequency or time resource block, we have factor $1/2$ for the average spectral efficiency in \eqref{e:tp def}. It can then be  derived by multiplying factor $1/2$ and replacing $\rho$ as $\rho=P/\sigma_{w}^2$ of \eqref{e:tp L1} and \eqref{e:tp L2}. 

With respect to the average spectral efficiency, we now discuss which transmission scheme is better for a given transmit power $P$. Since the direct comparison by using the spectral efficiency expressions in \eqref{e:tp L1} and \eqref{e:tp L2} for Scheme~1 and Scheme~2 does not yield any tractable analytical expressions for $P$, we compare their asymptotic expressions where the corresponding spectral efficiency expressions for Scheme~2 can be given with the aid of \eqref{e:tp L1 p asy} and \eqref{e:tp LL p asy} as 
\begin{align}\label{e:tp p asy 1-way}
\Rave_{\rm one}^{\infty} &  \xrightarrow{} \left\{
\begin{array}{ll}
\frac{\log (P)-\log \left(\frac{\sigma_w^2}{\sigma ^4}\right)-2 \epsilon}{2\log (2)};  &\text{ for }L=1  \\
\frac{\log (P)+2 \psi ^{(0)}(L k)-\log\left(\frac{\sigma_w^2}{\theta^2} \right)}{2\log (2)};   &\text{ for }L\geq 2 
\end{array}
\right..
\end{align}

Now we seek the condition for which Scheme~1 outperforms Scheme~2. 
\begin{lemma} \label{Lemma:prange}
The transmit power boundary where  Scheme~1 outperforms Scheme~2 can be approximately given  
\begin{itemize}
    \item 
for $\sigma_i^2=\omega$ as 
\begin{align}
P  & > \left\{
\begin{array}{ll}
\left(\frac{\omega+\sigma_w^2}{\sigma_w \sigma^2} \right)^2 {\e{2\epsilon}}; 
&\text{for~}L=1  \\
\left(\frac{\omega+\sigma_w^2}{\sigma_w \theta} \right)^2 {\e{-2\psi ^{(0)}(L k)}};
 & \text{for~}L\geq 2.
\end{array}
\right.
\end{align}
and 
   \item
for $\sigma_i^2=\omega P$ as
\begin{equation}
P  <\left\{
\begin{array}{ll}
{\e{2\log (2) \Rave_{L= 1}\left(\frac{1}{\omega}\right) + \log \left(\frac{\sigma_w^2}{\sigma ^4}\right)+2 \epsilon}};
\qquad\text{  ~}L=1  \\
{\e{2\log (2) \Rave_{L\geq 2}\left(\frac{1}{\omega}\right) + \log \left(\frac{\sigma_w^2}{\theta^2}\right) - 2 \psi ^{(0)}(L k)}};
  \text{}L\geq 2
\end{array}
\right.
\end{equation}
\end{itemize}
\end{lemma}

\begin{IEEEproof}
We can derive these with direct comparisons $\Rave_{\rm one}^{\infty}<\Rave_{\rm L=1}^{\infty}$ and $\Rave_{\rm one}^{\infty}<\Rave_{\rm L\geq 2}^{\infty}$ by using \eqref{e:tp L1 p asy},  \eqref{e:tp LL p asy} and \eqref{e:tp p asy 1-way}.  
\end{IEEEproof}
Since there are no simultaneous user transmissions, these analytical expressions are also valid for non-reciprocal channels.

\vspace{0mm}
\subsection{With Phase Adjustment Errors or Uncertainties}
For reciprocal channels, we assume that element $I_\ell$ introduces phase adjustment error $\epsilon_\ell$ due to channel estimation error or phase discretization error at RIS.  
 Under these scenarios, with the aid of \eqref{e:sinr u}, the equivalent instantaneous {\sf SINR} of each user can be written as 
\begin{align}\label{e_sinr_phase}
\gamma = \rho\left|\sum_{\ell=1}^{L}\alpha_\ell \beta_\ell \,\e{j \epsilon_\ell}\right|^2 \end{align} 
where $\epsilon_\ell$ is an i.i.d. RV which may be distributed as {\it uniform}, i.e., $\epsilon_\ell\sim \mathcal{U}(-\delta, \delta)$ and $f_{\epsilon_\ell}(t)  =\frac{1}{2\delta}$, or as the {\it von Mises}, i.e., $f_{\epsilon_\ell}(t)=\frac{\e{\kappa \cos(t-\mu)}}{2\pi I_0(\kappa)},\, t\in[0,2\pi]$ where $I_0(\kappa)$ is the modified Bessel function of order zero. The parameters $\mu$ which is a measure of location (the distribution is clustered around $\mu$) and $1/\kappa$ which is a measure of concentration  are analogous to $\mu$  and $\sigma^2$ (the mean and variance) in the normal distribution \cite{Badiu2020wcl}
\footnote{For non-reciprocal channels, the forward and reverse channels have different channel gains, and therefore,  we may not be able to completely compensate the phase of the received signal at each user. Therefore, the instantaneous {\sf SINR} of each user can also be written as in \eqref{e_sinr_phase}. Although the uniform and von Mises distributions may not be matched well with this case, we may still apply them as approximations with properly selected parameters: i) $\delta$ for the uniform distribution, and ii) $\mu$ and $\kappa$ for the von Mises distribution.}.
\begin{itemize}
    \item {\it For $\epsilon_\ell\sim \mathcal{U}(-\pi, \pi)$:} For the uniform distribution, this may be the worst case scenario.  Then,  the CDF of {\sf SINR} can be given as in the following lemma. 
    \begin{lemma}\label{thm-3}
For i.i.d. Rayleigh RVs $\alpha_\ell,\beta_\ell\sim \mathrm {Rayleigh} (\sigma/\sqrt{2} )$, and a uniformly distributed $\epsilon_\ell\sim \mathcal{U}(-\pi, \pi)$,  the CDF of $\gamma =\rho\left|\sum_{\ell=1}^L\alpha_\ell \beta_\ell \, \e{j\epsilon_\ell}\right|^2$ where $\rho>0$ can be given as 
\begin{equation}\label{e_cdf_snr_phase}
   F_\gamma(t)=1-\frac{2}{\Gamma(L)\sigma^{2L}} \left(\sqrt{\frac{t}{\rho }}\right)^{L} K_L\left(\frac{2 }{\sigma ^2}\sqrt{\frac{t}{\rho }}\right) 
\end{equation}
where $K_{n}\left(\cdot \right)$ is the modified Bessel function of the second kind of order $n$. \end{lemma}
\begin{IEEEproof}
By using the CDF of a cascade channel in \cite[eq.~(7)]{Haibin2014Elsevier} and a linear RV transformation, we can conclude the proof.
\end{IEEEproof}
Thus, the exact outage probability can be evaluated by using \eqref{e_cdf_snr_phase} as $P_{\rm out} = F_{\gamma}\left(\gamma_{\rm th}\right)$.\\ 
With the aid of \eqref{e:tp def}, \eqref{e:tp L1} and \eqref{e_cdf_snr_phase} , the exact average spectral efficiency can be evaluated as 
\begin{align}\label{e:tp phase error}
\Rave(\rho) & =  \frac{1}{\log(2)\Gamma(L)\sigma^{2L}\rho^{\frac{L}{2}}} {\sf G}_{1,3}^{3,1}\left(\frac{1}{\sigma^4 \rho }\Bigg|
\begin{array}{c}
 -\frac{L}{2} \\
 -\frac{L}{2},-\frac{L}{2},\frac{L}{2} \\
\end{array}
\right).
\end{align} 
\item {\it For $\epsilon_\ell\sim \mathcal{U}(-\delta, \delta)$ where $\delta\neq \pi$ or $\epsilon_\ell\sim$ von Mises distribution:} We can re-write \eqref{e_sinr_phase} as 
\begin{align}\label{e_sinr_phase2}
\gamma = \rho \left[\left(\sum_{\ell=1}^L \alpha_\ell \beta_\ell\cos \theta_\ell\right)^2 +\left(\sum_{\ell=1}^L \alpha_\ell \beta_\ell \sin \theta_\ell\right)^2\right] 
\end{align} 
where the derivation of $F_\gamma(t)$ may be difficult in these cases. One can approximate each sum by a gamma RV, and then use the bi-variate gamma distribution to derive an approximation for $F_\gamma(t)$. Since this is a challenging problem, we leave it as a future work.  
\end{itemize}

\textcolor{black}
{
For non-reciprocal channels, optimization based on CSI is involved. In this case, two models are usually utilized to describe the CSI measurement error \cite{zhou2020framework,zhou2019robust}.
In the first model, the CSI error of the concatenated channel from user $U_1$ (or $U_2$) to user $U_2$ (or $U_1$) through the RIS is bounded, which indicates the Frobenius norm of the cascaded channel is upper limited. By resorting to the methods in  \cite{zhou2020framework,zhou2019robust}, a combination of S-procedure and penalty convex-concave procedure (CCP) can help to find the phase setting at the RIS to realize a given minimum user SINR.
In the second model, the CSI error of the concatenated channel from user $U_1$ (or $U_2$) to user $U_2$ (or $U_1$) through the RIS is subject to Gaussian random distribution, rate outage  probability should be imposed on every link in this case.
By utilizing the methods in \cite{zhou2020framework}, including Bernstein-type inequality and penalty CCP, we are also able to find the phase setting at the RIS to realize a given minimum user SINR.
Then bisection search of the minimum user SINR can be adopted to find the maximal achievable one under both of the above two models.
Due to the limit of space, details are not presented here. The interested reader may refer to \cite{zhou2020framework,zhou2019robust}.
}

\vspace{-0mm}
\section{Simulation  Results} \label{s:num}

In this section, we investigate the performance of the RIS-aided two--way networks. We set channel variance $\sigma^2=1$. 
Since the thermal noise floor for 1\,Hz bandwidth at room temperature can even be -174\,dBm, we use -70\,dBm to represent a more noisy scenario. 
All presented illustrations include average results over $10^6$ and $10^3$ independent channel realizations for the outage probability and the average spectral efficiency calculations, respectively. 
\vspace{0mm}
\subsection{For $L=1$ with Reciprocal Channels}
\begin{figure*}
	\centering
	\subfloat[The outage probability vs $P$ .]{
		\label{f_outL1}
		\includegraphics[width=0.5\textwidth]{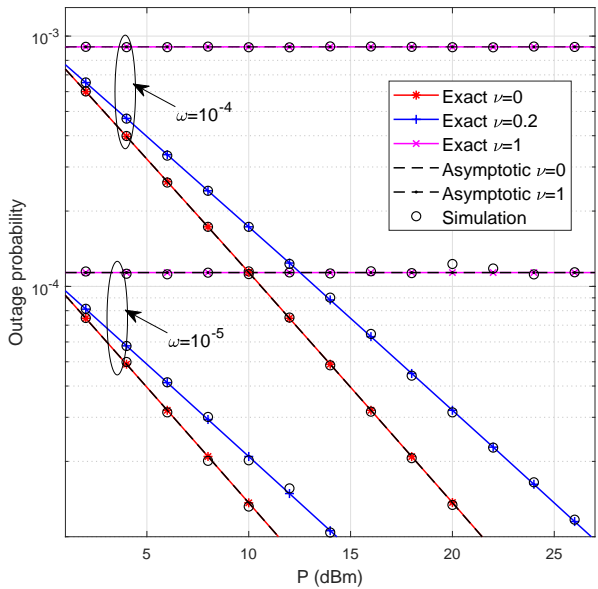}} 
        	\subfloat[The  spectral efficiency  vs $P$.]{
		\label{f_tpL1}
		\includegraphics[width=0.5\textwidth]{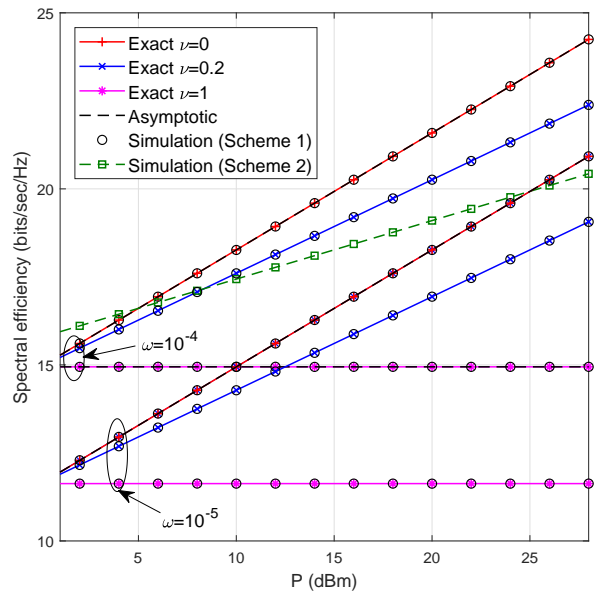}}
	\caption{Performance of reciprocal channels when $L=1$ for different loop-interference $\sigma_i^2-\omega P^\nu$.\vspace{-0mm}}
	\label{f_L1}
\end{figure*}
Fig.~\ref{f_L1} shows the outage probability and average spectral efficiency vs $P$ for $L=1$. 
Several observations are gained: 
i) Our analytical results in \eqref{e:out L1} and \eqref{e:tp L1} exactly match with the simulation results, which confirms the accuracy of our analysis; 
ii) For different loop interference $\sigma_i^2-\omega P^\nu$, we notice that the outage decreases at a rate of $\log(P)/P$ and the spectral efficiency increases at a rate of $\log(P)$ when $\nu=0$,  and both have floors when $\nu=1$ due to the transmit-power dependent interference. These have been analytically proved in \eqref{e:out L1 p asy} and \eqref{e:tp L1 p asy}. As we expect, when $\nu\in(0,1)$, e.g., $\nu=0.2$, the outage and spectral efficiency are in between $\nu=0$ and $\nu=1$ cases; 
iii) When $\omega$ reduces from $10^{-4}$ to $10^{-5}$, the outage and spectral efficiency improve around 9\,dB and 3.32\,[bits/sec/Hz], respectively, for each case; and 
iv) two--way communications with Scheme~2 outperforms Scheme~1 when $P<5$\,dBm and $P<25$\,dBm for $\omega=10^{-5}$ and $\omega=10^{-4}$, respectively, with $\nu=0$. For $\nu=1$, Scheme~2 outperforms Scheme~1 in the entire simulated region\footnote{We do not show the outage probability of Scheme~2 because it always outperforms Scheme~1 as long as the loop interference is non-zero.}. Therefore, it is important to keep the effect of loop interference independent of transmit power if two--way communications use Scheme~1.

\vspace{0mm}
\subsection{For $L\geq 2$ with Reciprocal Channels}

\begin{figure}
  \centering
  \includegraphics[width=0.5\textwidth]{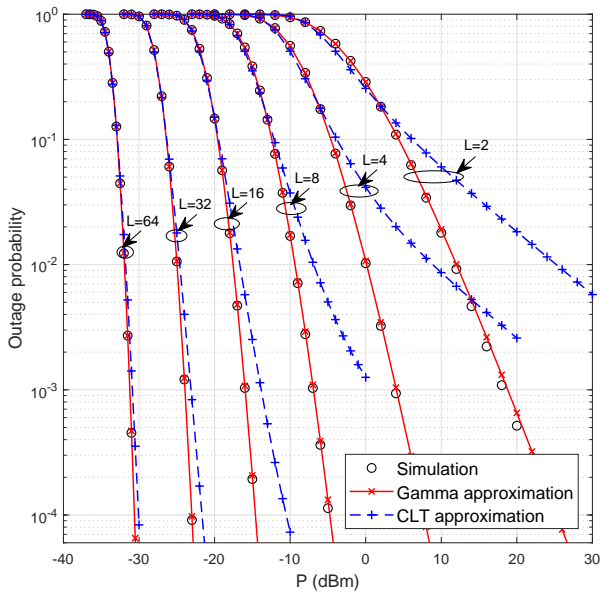}\\
  \caption{The outage probability vs $P$ when $\sigma_i^2 =10^{-4}$.\vspace{-0mm}
  }\label{f_outL2w}
\end{figure}

For $L\geq 2$, Fig.~\ref{f_outL2w} shows the outage probability vs $P$, when loop-interference is independent of transmit power $P$, i.e., $\sigma_i^2=\omega$.  
For a given $L$, the outage probability decreases with $\left[\log(P)/P\right]^L$ which confirms Lemma~\ref{Lemma:outage}. Although the outage probability decreases with $L$, the diminishing rate also decreases, as discussed earlier with respect to \eqref{e_Pdiff}. For example, when we increase $L$ from 2 to 4, we can save power around 14\,dBm at $10^{-3}$ outage. However, for the same outage, we can only save  power around 8\,dBm when we increase $L$ from 32 to 64. Interestingly, this figure confirms the accuracy of our gamma approximation. Moreover, it is more accurate than the CLT approximation even for $L=32$ or $L=64$.

\begin{figure*}
	\centering
		\subfloat[The  spectral efficiency vs $P$ when $\sigma_i^2=\omega=10^{-4}$.]{
		\label{f_tpL2w}
		\includegraphics[width=0.5\textwidth]{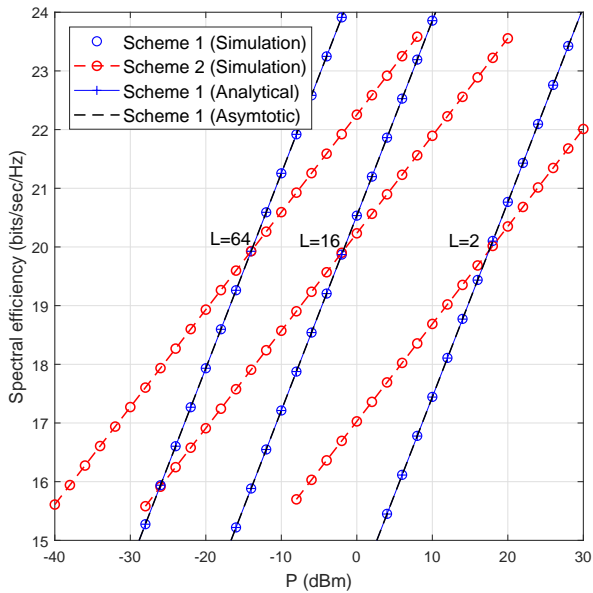}} 
	\subfloat[The  spectral efficiency vs $P$ when $\sigma_i^2=\omega P=10^{-4}P$.]{
		\label{f_tpL2wp}
		\includegraphics[width=0.5\textwidth]{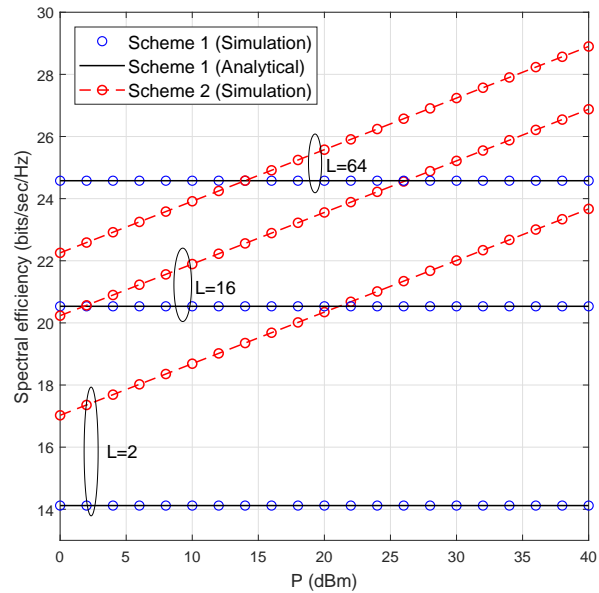}}
	\caption{The performance for reciprocal channels with $L\geq 2$.\vspace{-0mm}}
	\label{f_L2}
\end{figure*}

For $L\geq 2$, Fig.~\ref{f_L2} shows the  spectral efficiency vs $P$, when loop-interference is independent of transmit power $P$, i.e., $\sigma_i^2=\omega$, and linearly dependent of transmit power $P$, i.e., $\sigma_i^2=\omega P$. 
For any $L$, as shown in Fig.~\ref{f_tpL2w} and \eqref{e:tp LL p asy}, the average spectral efficiency increases in order of $\log(P)$ when  $\sigma_i^2=\omega$, which confirms Lemma~\ref{Lemma:tp}. According to the figure and \eqref{e_Pdiff}, while transmit power reduces by around 19\,dBm when $L$ increases from 2 to 16, we can only save 12\,dBm when $L$ increases from 16 to 64.  
We also plot the spectral efficiency of Scheme~1 and Scheme~2 in Fig.~\ref{f_tpL2w} where Scheme~1 starts to outperform Scheme~2 when $P$ increases where transition happens at $P\approx 17.5,~-1.8,-14.0$\,dBm for $L=2,16,64$, respectively. This compliments Lemma~\ref{Lemma:prange}. 
Fig.~\ref{f_tpL2wp} is for $\sigma_i^2=\omega P$ where we have spectral efficiency floors because loop-interference enhances with transmit power in Scheme~1. Due to this reason, as shown in the figure, Scheme~2  outperforms Scheme~1 when $P$ increases.   

\begin{figure*}
	\centering
	\subfloat[The transmit power vs $\omega$ of $\sigma_i^2=\omega$.]{
		\label{f_plim0}
		\includegraphics[width=0.5\textwidth]{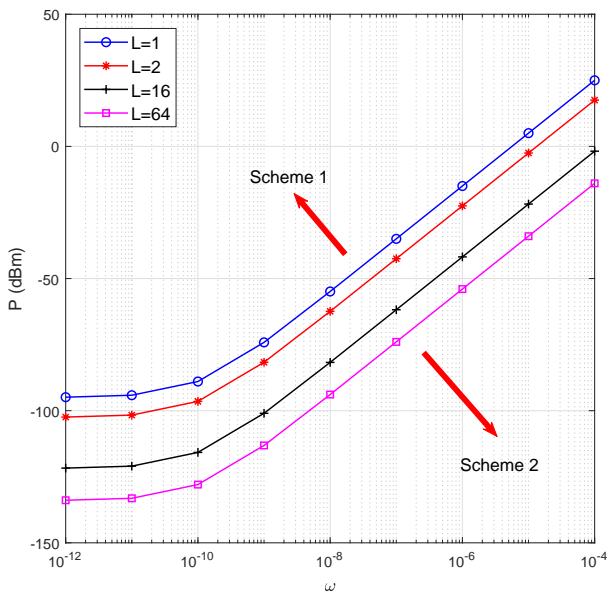}} 
        	\subfloat[The transmit power vs $\omega$ of $\sigma_i^2=\omega P$.]{
		\label{f_plim1}
		\includegraphics[width=0.5\textwidth]{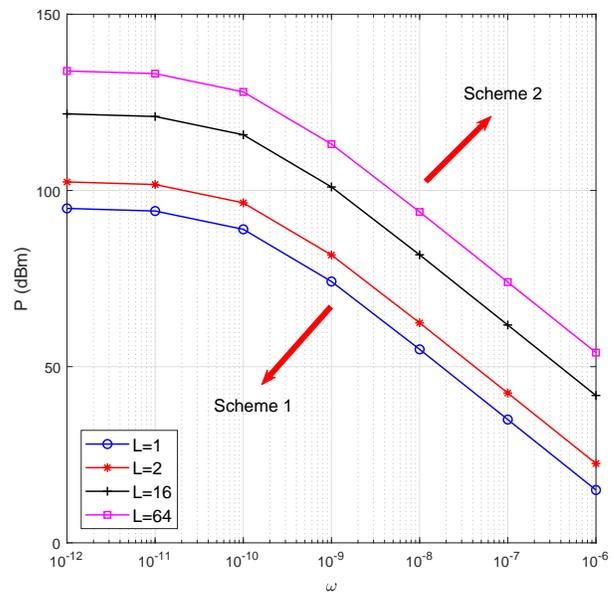}}
	\caption{The boundary of transmit power for Scheme~1 and Scheme~2.\vspace{-5mm}}
	\label{f_plim}
\end{figure*}

Fig.~\ref{f_plim} shows the transition boundary of transmit power $P$ where Scheme~1 outperforms Scheme~2 or vice versa. 
Based on the results in Lemma~\ref{Lemma:prange} and simulations, we plot $P$  vs $\omega$
for both cases $\sigma_i^2=\omega$ and $\sigma_i^2=\omega P$ in Fig.~\ref{f_plim0} and Fig.~\ref{f_plim1}, respectively. For  $\sigma_i^2=\omega$, Scheme~1 outperforms at high $P$, and the $P$ decreases when $L$ increases for given $\omega$. 
We have opposite observation for the other case $\sigma_i^2=\omega P$.  Moreover, when loop interference power is less than the noise power, i.e., $\omega<10^{-10}$, the noise power dominates, and we have power floor. For example, the power floor is around -120\,dBm with $L=16$ for  $\sigma_i^2=\omega$ case. 
\vspace{0mm}

\subsection{For Phase Adjustment Errors}

\begin{figure*}
	\centering
	\subfloat[The outage probability vs $P$ for $L=4,16$.]{
		\label{f_outphaserr}
		\includegraphics[width=0.5\textwidth]{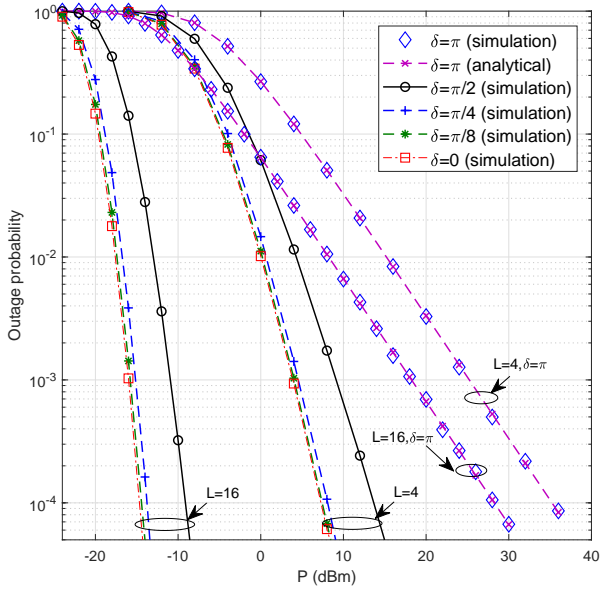}} 
        	\subfloat[The spectral efficiency vs $P$ for $L=4,32$.]{
		\label{f_tpphaserr}
		\includegraphics[width=0.5\textwidth]{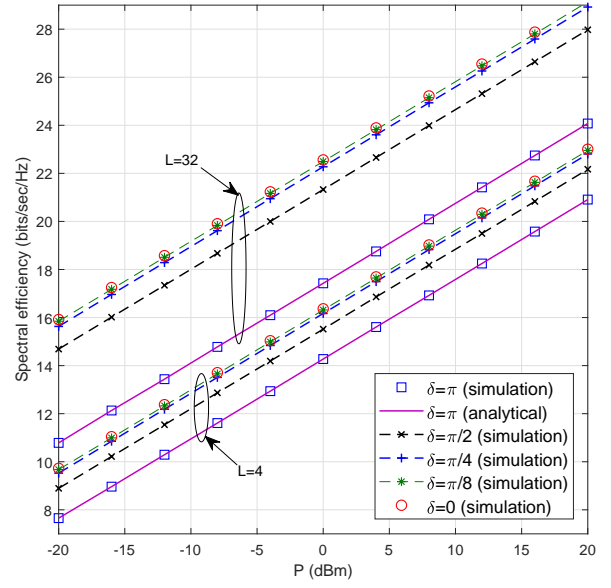}}
	\caption{The performance of reciprocal channels with Scheme~1 when phase adjustments at the RIS have errors $\epsilon_\ell,\forall \ell$, where $\epsilon_\ell\sim \mathcal{U}[-\delta, \delta]$ and $\delta=0,\pi/8,\pi/4,\pi/2$, or $\pi$.\vspace{-5mm}}
	\label{f_phserr}
\end{figure*}

Fig.~\ref{f_phserr} shows the outage probability and average spectral efficiency vs $P$ for different $L$ when there exists a phase adjustment error or uncertainty $\epsilon_\ell$ at the $\ell$th element of RIS. We assume that $\epsilon_\ell$ is an i.i.d. uniformly distributed RV as $\epsilon_\ell\sim \mathcal{U}[-\delta, \delta]$, where $\delta=0$ and $\delta=\pi$ represent no phase adjustment error (our main results of this paper) and random phase adjustment (the worst case scenario), respectively. we also assume that $\sigma_i^2=\omega=10^{-4}$. 
Several observations are made: 
i) Our analytical results for $\delta=\pi$ in \eqref{e_cdf_snr_phase} and \eqref{e:tp phase error} exactly match with the simulation results, which confirms the accuracy of our analysis; 
ii) For different $\delta$, we notice that the outage decreases when $\delta$ decreases where diversity gain changes from $\log(P)/P$ at $\delta\xrightarrow{} \pi$ to  $\left(\log(P)/P\right)^L$ at $\delta\xrightarrow{} 0$. Moreover, we achieve $10^{-4}$ outage probability with $L=16$ at $P\approx 27$\,dBm for $\delta=\pi$, at $P\approx -9$\,dBm for $\delta=\pi/2$, and at $P\approx -14$\,dBm for $\delta=\pi/4,\pi/8,\text{ or }0$ where, compared to $\delta=\pi$, we save $99.97$\% or $99.99$\% of power when $\delta=\pi/2$ or $\delta\leq \pi/4$, respectively;   
iii) For different $\delta$, we also notice that the spectral efficiency increases at a rate of $\log(P)$ for any $\delta$. Further, we gain $22.4\%, 27.8\%, 29.1\%$ and $29.5\%$ of spectral efficiency over $\delta=\pi$ when  $\delta=\pi/2,\pi/4,\pi/8$, and $0$, respectively; and 
iv) We have a negligible performance gap with no error case (i.e., $\delta=0$) when $\delta\leq \pi/8$.

\subsection{For Non-reciprocal Channels}
\begin{figure*}
	\centering
	\subfloat[The spectral efficiency vs $P$ for $L=8$.]{
		\label{f_nonrecL8}
		\includegraphics[width=0.5\textwidth]{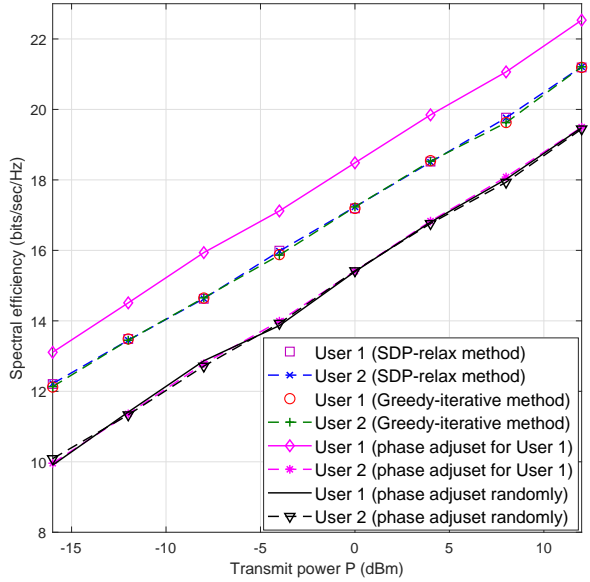}} 
        	\subfloat[The spectral efficiency vs $P$ for $L=1,2,4,16$.]{
		\label{f_nonrecLall}
		\includegraphics[width=0.5\textwidth]{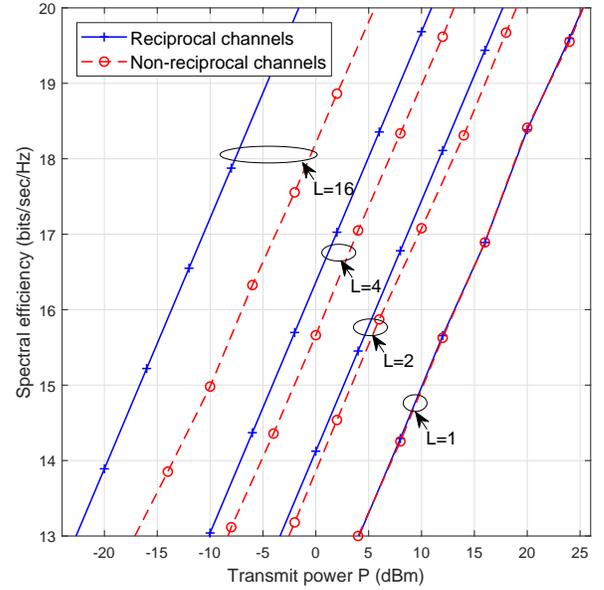}}
	\caption{The performance of non-reciprocal channels with Scheme~1.\vspace{-0mm}}
	\label{f_nonrec}
\end{figure*}

Fig.~\ref{f_nonrecL8} plots the spectral efficiency vs transmit power of both users for three types of phase adjustment techniques: 1) two fairness algorithms (SDP-relax method and greedy-iterative method); 2) phase is adjusted based on $U_1$
, i.e., $\phi_\ell=\varphi_{r,\ell}+\psi_{t,\ell}$; and 3)  phase is randomly adjusted. Both users $U_1$ and $U_2$ have the almost same spectral efficiency with both fairness methods, which confirms the user fairness and also corroborates that both methods provide almost same numerical values. When phase is adjusted based on $U_1$, $U_1$ has the best performance among all, and it has around 7.5\% spectral efficiency improvement at $P=0$\,dBm compared to  two fairness algorithms. However, $U_2$ has the worst performance which is very similar to the case of random phase adjustment where both users have similar poor performance. The spectral efficiency reduction is around 10.5\% at $P=0$\,dBm compared to the fairness algorithms.     

Fig.~\ref{f_nonrecLall} plots the spectral efficiency vs transmit power of $U_1$ for reciprocal and non-reciprocal channels with different $L$. When $L=1$, both cases show the same spectral efficiency as phase adjustment does not effect the performance. However, when $L\geq 2$, the reciprocal channel case outperforms the non-reciprocal channel case. The reason is that, for each reflective element with reciprocal channels, the effective phase for the {\sf SINR} is common for both users and the corresponding optimum phase can also maximize the each user {\sf SINR}. However, for each reflective element with non-reciprocal channels,  the effective phases for the {\sf SINR}s of two users are different and the corresponding optimum phases which maximize the minimum user {\sf SINR} do not maximize the each user {\sf SINR}. Therefore, we lose some spectral efficiency compared with reciprocal channel case. As illustrated in Fig.~\ref{f_nonrecLall}, the spectral efficiency gap between these two cases increases when $L$ increases, e.g., the difference between transmit powers which achieve spectral efficiency 15\, [bits/sec/Hz] are 0.3, 2.0 and  6.5\,dBm for $L=2,4$ and 16, respectively.

\vspace{0mm}
\section{Conclusion}

In this work, RIS assisted systems have been proposed  for  two--way  wireless communications.  
Two possible transmission schemes are introduced where Scheme~1 and Scheme~2 require one and two resource blocks (time or frequency), respectively.  
For both reciprocal and non-reciprocal channels, Scheme~1 is the main focus of this work. For the optimal phases of the RIS elements over reciprocal channels, the exact {\sf SINR} outage probability and average spectral efficiency have been derived for a single-element RIS. Since the exact performance analysis for a multiple-element RIS seems intractable, approximations have been derived for the outage probability and average spectral efficiency. In this respect, a product of two Rayleigh random variables  approximated by a gamma random variable. 
Moreover, asymptotic analysis has been conducted for high {\sf SINR} $\rho$ regime. Our analysis reveals  that the outage probability decreases at the rate of $\left( \log(\rho)/\rho\right)^L$, whereas spectral efficiency increases at the rate of $\log(\rho)$. Moreover, we  observe either an  outage or spectral efficiency floor caused by transmit power dependent loop interference. 
Cross over boundary, where Scheme~1 outperforms Scheme~2 and vice versa, has also been approximately derived based on the asymptotic results. 
For non-reciprocal channels, an optimization problem is formulated, which optimizes the phases of RIS elements so as to maximize the minimum user {\sf SINR}. Although being non-convex, sub-optimal solution is found by relaxing and then transforming the original optimization problem to be a SDP problem for multiple-element RIS and closed-form solution is found for single-element RIS. 
Simulation results have illustrated that the rate of spectral efficiency increment 
or transmit power saving reduces when number of elements increases. A network with reciprocal channels outperforms in terms of outage or spectral efficiency the same with non-reciprocal channels.  


\begin{appendices}
		
\section{Proof of Theorem~\ref{thm-1}}\label{App:thm-1}
With the aid of asymptotic expansion of ${\sf K}_1\left(x\right)$ at $x\approx 0$ \cite[eq.~8.446]{gradshteyn2007book}, we have, for $a>0$, 
\[{\sf K}_1\left(a\sqrt{x}\right)\xrightarrow{}\frac{1}{a \sqrt{x}}+\frac{\sqrt{x}}{4}  \left(a \log (x)+2 \epsilon  a-a+2 a \log \left(\frac{a}{2}\right)\right).\]

For $L=1$, since the outage expression in \eqref{e:out L1} contains the term $a\sqrt{x}{\sf K}_1\left(a\sqrt{x}\right)$ where $x=1/\rho$ and $a=\frac{2}{\sigma^2} \sqrt{\gamma_{\rm th}}$, considering the dominant terms, we have a high {\sf SINR} approximation as $P_{\rm out|L= 1}^{\infty}(\gamma_{\rm th})  \xrightarrow{}
\frac{\gamma_{\rm th}}{\sigma ^4} \frac{\log (\rho)}{\rho}$. 

For $L\geq 2$, we have a bound as $\PR\left(\max_{\ell\in[1,L]}\zeta_\ell\leq \sqrt{\frac{\gamma_{\rm th}}{\rho L^2}}\right)\leq  \PR\left(\sum_{\ell=1}^{L}\zeta_\ell\leq \sqrt{\frac{\gamma_{\rm th}}{\rho}}\right)\leq \prod_{\ell=1}^{L}\PR\left(\zeta_\ell\leq \sqrt{\frac{\gamma_{\rm th}}{\rho}}\right)$
from which we can write 
$F_{\zeta_\ell}\left(\sqrt{\frac{\gamma_{\rm th}}{\rho L^2}}\right)^L\leq  \PR\left(\sum_{\ell=1}^{L}\zeta_\ell\leq \sqrt{\frac{\gamma_{\rm th}}{\rho}}\right)\leq F_{\zeta_\ell}\left(\sqrt{\frac{\gamma_{\rm th}}{\rho}}\right)^L.$ 
This can be written with outage probabilities as $\left[ P_{\rm out|L=1}\left(\frac{\gamma_{\rm th}}{L^2}\right)\right]^L\leq  P_{\rm out|L\geq 2}(\gamma_{\rm th}) \leq \left[P_{\rm out|L=1}(\gamma_{\rm th}) \right]^L.$
We have 
$\left[\frac{\gamma_{\rm th}}{\sigma ^4 L^2} \frac{\log (\rho)}{\rho}\right]^L\leq  P_{\rm out|L\geq 2}(\gamma_{\rm th}) \leq \left[\frac{\gamma_{\rm th}}{\sigma ^4} \frac{\log (\rho)}{\rho} \right]^L$  
and proves the theorem. 

\section{Proof of Theorem~\ref{thm-2}}\label{App:thm-2}
We first find the Mellin Transform of ${\sf G}_{1,3}^{3,1}\left(\cdot\right)$ in \eqref{e:tp L1} 
by using \cite{Mathematica}, which gives $-1/2 \pi  (2 s-1) a^{-s} \sec (\pi  s) \Gamma \left(s-1/2\right)^2$ where $a=1/(\sigma^4 \rho)$ and this transform exists within the residue $1/2 < {\sf Re}[s] < 3/2$. We now sum to the left of the strip starting with $s=1/2$ which results
\[{\sf G}_{1,3}^{3,1}\left(a x\left|
\begin{array}{c}
 -\frac{1}{2} \\
 -\frac{1}{2},-\frac{1}{2},\frac{1}{2} \\
\end{array}
\right.\right)\xrightarrow{} \frac{-\log (ax)-2 \epsilon }{\sqrt{ax}}.\]
Then, for $L=1$, the average spectral efficiency expression in \eqref{e:tp L1} can be approximated at high {\sf SINR}, i.e., $\rho\gg 1$, as 
$\Rave_{\rm L=1}^{\infty}  \xrightarrow{}
\frac{\log (\rho )-\log \left(\frac{1}{\sigma ^4}\right)-2 \epsilon}{\log (2)}$. 
For $L\geq 2$, with the aid of \eqref{e:tp L2}, since the terms associated with hypergeometric functions have negligible effect at $\rho\gg 1$, the average spectral efficiency expression can be approximated as $\Rave_{\rm L\geq 2}^{\infty} \xrightarrow{} \frac{\log (\rho )+ 2 \log (\theta )+2 \psi ^{(0)}(L k)  }{\log(2)}.$ 
These asymptotic expressions 
increase at rate $\log (\rho )$ as $\rho$ increases, which proves the theorem. 
			
\end{appendices}


\vspace{-5mm}
\bibliographystyle{IEEEtran}
\bibliography{reference1,IEEEabrv}

\begin{thebibliography}{10}
\providecommand{\url}[1]{#1}
\csname url@samestyle\endcsname
\providecommand{\newblock}{\relax}
\providecommand{\bibinfo}[2]{#2}
\providecommand{\BIBentrySTDinterwordspacing}{\spaceskip=0pt\relax}
\providecommand{\BIBentryALTinterwordstretchfactor}{4}
\providecommand{\BIBentryALTinterwordspacing}{\spaceskip=\fontdimen2\font plus
\BIBentryALTinterwordstretchfactor\fontdimen3\font minus
  \fontdimen4\font\relax}
\providecommand{\BIBforeignlanguage}[2]{{%
\expandafter\ifx\csname l@#1\endcsname\relax
\typeout{** WARNING: IEEEtran.bst: No hyphenation pattern has been}%
\typeout{** loaded for the language `#1'. Using the pattern for}%
\typeout{** the default language instead.}%
\else
\language=\csname l@#1\endcsname
\fi
#2}}
\providecommand{\BIBdecl}{\relax}
\BIBdecl

\bibitem{atapattu2020wcnc}
S.~Atapattu, R.~Fan, P.~Dharmawansa, G.~Wang, and J.~Evans, ``Two-way
  communications via reconfigurable intelligent surface,'' in \emph{{IEEE}
  Wireless Commun. and Networking Conf. (WCNC)}, 2020, {A}ccepted.

\bibitem{renzo2019corr}
M.~D. Renzo \emph{et~al.}, ``Smart radio environments empowered by {AI}
  reconfigurable meta-surfaces: An idea whose time has come,'' \emph{EURASIP J.
  Wireless Commun. Netw.}, vol. 2019:129, May 2019.

\bibitem{He2019wcoml}
Z.~{He} and X.~{Yuan}, ``Cascaded channel estimation for large intelligent
  metasurface assisted massive {MIMO},'' \emph{IEEE Wireless Commun. Lett.},
  vol.~9, no.~2, pp. 210--214, 2020.

\bibitem{Zheng2020coml}
B.~{Zheng}, Q.~{Wu}, and R.~{Zhang}, ``Intelligent reflecting surface-assisted
  multiple access with user pairing: {NOMA} or {OMA}?'' \emph{{IEEE} Commun.
  Lett.}, vol.~24, no.~4, pp. 753--757, 2020.

\bibitem{Zhao2020coml}
W.~{Zhao}, G.~{Wang}, S.~{Atapattu}, T.~A. {Tsiftsis}, and C.~{Tellambura},
  ``Is backscatter link stronger than direct link in reconfigurable intelligent
  surface-assisted system?'' \emph{{IEEE} Commun. Lett.}, 2020, {A}ccepted.

\bibitem{Wu2018gcom}
Q.~{Wu} and R.~{Zhang}, ``Intelligent reflecting surface enhanced wireless
  network: Joint active and passive beamforming design,'' in \emph{{IEEE}
  Global Telecommn. Conf. (GLOBECOM)}, Dec. 2018.

\bibitem{Yu2019iccc}
X.~{Yu}, D.~{Xu}, and R.~{Schober}, ``{MISO} wireless communication systems via
  intelligent reflecting surfaces : (invited paper),'' in \emph{IEEE/CIC Int.
  Conf. Commun. in China (ICCC)}, 2019, pp. 735--740.

\bibitem{Abeywickrama2019arx}
S.~Abeywickrama, R.~Zhang, and C.~Yuen, ``Intelligent reflecting surface:
  Practical phase shift model and beamforming optimization,'' Available:
  https://arxiv.org/abs/1907.06002.

\bibitem{Huang2019twc}
C.~{Huang}, A.~{Zappone}, G.~C. {Alexandropoulos}, M.~{Debbah}, and C.~{Yuen},
  ``Reconfigurable intelligent surfaces for energy efficiency in wireless
  communication,'' \emph{{IEEE} Trans. Wireless Commun.}, vol.~18, no.~8, pp.
  4157--4170, Aug. 2019.

\bibitem{Guo2019gcom}
H.~{Guo}, Y.~{Liang}, J.~{Chen}, and E.~G. {Larsson}, ``Weighted sum-rate
  maximization for intelligent reflecting surface enhanced wireless networks,''
  in \emph{{IEEE} Global Telecommn. Conf. (GLOBECOM)}, 2019.

\bibitem{Cui2019wcl}
M.~{Cui}, G.~{Zhang}, and R.~{Zhang}, ``Secure wireless communication via
  intelligent reflecting surface,'' \emph{IEEE Wireless Commun. Lett.}, vol.~8,
  no.~5, pp. 1410--1414, Oct. 2019.

\bibitem{Shen2019coml}
H.~{Shen}, W.~{Xu}, W.~{Xu}, S.~{Gong}, Z.~{He}, and C.~{Zhao}, ``Secrecy rate
  maximization for intelligent reflecting surface assisted multi-antenna
  communications,'' \emph{IEEE Commun. Lett.}, vol.~23, no.~9, pp. 1488--1492,
  Sep. 2019.

\bibitem{Di2020tvt}
B.~{Di}, H.~{Zhang}, L.~{Li}, L.~{Song}, Y.~{Li}, and Z.~{Han}, ``Practical
  hybrid beamforming with limited-resolution phase shifters for reconfigurable
  intelligent surface based multi-user communications,'' \emph{{IEEE} Trans.
  Veh. Technol.}, pp. 1--1, 2020.

\bibitem{Wu2019icassp}
Q.~{Wu} and R.~{Zhang}, ``Beamforming optimization for intelligent reflecting
  surface with discrete phase shifts,'' in \emph{Proc. {IEEE} Int. Conf.
  Acoustics, Speech, and Signal Processing (ICASSP)}, 2019, pp. 7830--7833.

\bibitem{Huang2020jsac}
C.~Huang, , R.~Mo, and C.~Yuen, ``Reconfigurable intelligent surface assisted
  multiuser {MISO} systems exploiting deep reinforcement learning,''
  \emph{{IEEE} J. Select. Areas Commun.}, 2020, {A}ccepted.

\bibitem{Han2019tvt}
Y.~{Han}, W.~{Tang}, S.~{Jin}, C.~{Wen}, and X.~{Ma}, ``Large intelligent
  surface-assisted wireless communication exploiting statistical {CSI},''
  \emph{{IEEE} Trans. Veh. Technol.}, vol.~68, no.~8, pp. 8238--8242, Aug.
  2019.

\bibitem{Nadeem2020twc}
Q.~{Nadeem}, A.~{Kammoun}, A.~{Chaaban}, M.~{Debbah}, and M.~{Alouini},
  ``Asymptotic max-min {SINR} analysis of reconfigurable intelligent surface
  assisted {MISO} systems,'' \emph{{IEEE} Trans. Wireless Commun.}, 2020.

\bibitem{Jung2019arx}
M.~Jung, W.~Saad, Y.~Jang, G.~Kong, and S.~Choi, ``Performance analysis of
  large intelligent surfaces ({LIS}s): Asymptotic data rate and channel
  hardening effects,'' Available: https://arxiv.org/abs/1810.05667.

\bibitem{Basar2019acc}
E.~{Basar}, M.~{Di Renzo}, J.~{De Rosny}, M.~{Debbah}, M.~{Alouini}, and
  R.~{Zhang}, ``Wireless communications through reconfigurable intelligent
  surfaces,'' \emph{IEEE Access}, vol.~7, pp. 116\,753--116\,773, 2019.

\bibitem{Badiu2020coml}
M.~{Badiu} and J.~P. {Coon}, ``Communication through a large reflecting surface
  with phase errors,'' \emph{IEEE Wireless Commun. Lett.}, vol.~9, no.~2, pp.
  184--188, 2020.

\bibitem{Zhao2020wcoml}
W.~{Zhao}, G.~{Wang}, S.~{Atapattu}, T.~A. {Tsiftsis}, and X.~{Ma},
  ``Performance analysis of large intelligent surface aided backscatter
  communication systems,'' \emph{IEEE Wireless Commun. Lett.}, 2020,
  {A}ccepted.

\bibitem{Atapattu2013tcom}
S.~{Atapattu}, Y.~{Jing}, H.~{Jiang}, and C.~{Tellambura}, ``Relay selection
  schemes and performance analysis approximations for two-way networks,''
  \emph{{IEEE} Trans. Commun.}, vol.~61, no.~3, pp. 987--998, Mar. 2013.

\bibitem{Hanzo2016proc}
Z.~Zhang, K.~Long, A.~V. Vasilakos, and L.~Hanzo, ``Full-duplex wireless
  communications: Challenges, solutions, and future research directions,''
  \emph{Proc. {IEEE}}, vol. 104, no.~7, pp. 1369--1409, Jul. 2016.

\bibitem{Atapattu2010icst}
S.~{Atapattu}, Y.~{Jing}, H.~{Jiang}, and C.~{Tellambura}, ``Opportunistic
  relaying in two-way networks (invited paper),'' in \emph{5th Int. ICST Conf.
  Commun. and Networking in China}, 2010, pp. 1--8.

\bibitem{Atapattu2019tcom}
S.~{Atapattu}, P.~{Dharmawansa}, M.~{Di Renzo}, C.~{Tellambura}, and J.~S.
  {Evans}, ``Multi-user relay selection for full-duplex radio,'' \emph{{IEEE}
  Trans. Commun.}, vol.~67, no.~2, pp. 955--972, 2019.

\bibitem{Zheng2019wcl}
B.~{Zheng} and R.~{Zhang}, ``Intelligent reflecting surface-enhanced {OFDM}:
  Channel estimation and reflection optimization,'' \emph{IEEE Wireless Commun.
  Lett.}, vol.~9, no.~4, pp. 518--522, 2020.

\bibitem{gradshteyn2007book}
I.~S. Gradshteyn and I.~M. Ryzhik, \emph{Table of Integrals, Series and
  Products}, 7th~ed.\hskip 1em plus 0.5em minus 0.4em\relax Academic Press Inc,
  2007.

\bibitem{Atapattu2011twc}
S.~{Atapattu}, C.~{Tellambura}, and H.~{Jiang}, ``A mixture gamma distribution
  to model the {SNR} of wireless channels,'' \emph{{IEEE} Trans. Wireless
  Commun.}, vol.~10, no.~12, pp. 4193--4203, Dec. 2011.

\bibitem{Prudnikov83book}
A.~P. Prudnikov, Y.~A. Bry\v{c}kov, and O.~I. Mari\v{c}ev, \emph{Integrals and
  Series of Special Functions}.\hskip 1em plus 0.5em minus 0.4em\relax Moscow,
  Russia, Russia: Science, 1983.

\bibitem{Annamalai2009ccnc}
A.~{Annamalai}, C.~{Tellambura}, and J.~{Matyjas}, ``A new twist on the
  generalized {M}arcum {Q}-function ${Q}_{M}(a, b)$ with fractional-order ${M}$
  and its applications,'' in \emph{IEEE Consumer Commun. Networking Conf.
  (CCNC)}, Jan. 2009.

\bibitem{luo2010semidefinite}
Z.-Q. Luo, W.-K. Ma, A.~M.-C. So, Y.~Ye, and S.~Zhang, ``Semidefinite
  relaxation of quadratic optimization problems,'' \emph{IEEE Signal Proc.
  Mag.}, vol.~27, no.~3, pp. 20--34, May 2010.

\bibitem{cvx}
M.~Grant and S.~Boyd, ``{CVX}: Matlab software for disciplined convex
  programming, version 2.1,'' \url{http://cvxr.com/cvx}, Dec. 2018.

\bibitem{Boyd}
S.~Boyd and L.~Vandenberghe, \emph{Convex Optimization}.\hskip 1em plus 0.5em
  minus 0.4em\relax Cambridge University Press, 2004.

\bibitem{Badiu2020wcl}
M.~{Badiu} and J.~P. {Coon}, ``Communication through a large reflecting surface
  with phase errors,'' \emph{{IEEE} Wireless Commun. Lett.}, vol.~9, no.~2, pp.
  184--188, Feb. 2020.

\bibitem{Haibin2014Elsevier}
H.~Liu, H.~Ding, L.~Xiang, J.~Yuan, and L.~Zheng, ``Outage and {BER}
  performance analysis of cascade channel in relay networks.'' in
  \emph{Elsevier, Procedia Computer Science}, vol.~34, 2014, pp. 23--30.

\bibitem{zhou2020framework}
G.~Zhou, C.~Pan, H.~Ren, K.~Wang, and A.~Nallanathan, ``A framework of robust
  transmission design for {IRS}-aided {MISO} communications with imperfect
  cascaded channels,'' \emph{arXiv preprint arXiv:2001.07054}, 2020.

\bibitem{zhou2019robust}
G.~Zhou, C.~Pan, H.~Ren, K.~Wang, M.~Di~Renzo, and A.~Nallanathan, ``Robust
  beamforming design for intelligent reflecting surface aided {MISO}
  communication systems,'' \emph{arXiv preprint arXiv:1911.06237}, 2019.

\bibitem{Mathematica}
``Mathematica, {V}ersion 12.0,'' \url{http://functions.wolfram.com/
  HypergeometricFunctions/MeijerG/06/01/03/01/0003/}, accessed: 2019-12-30.

\end{thebibliography}

\end{document}